\documentclass[conference]{IEEEtran}
\IEEEoverridecommandlockouts
\pdfoutput=1
\usepackage{cite}
\usepackage{amsmath,amssymb,amsfonts}
\usepackage{algorithmic}
\usepackage{graphicx}
\usepackage{textcomp}
\usepackage{xcolor}

\usepackage{algorithm}
\usepackage{multirow}
\usepackage{multicol}
\usepackage{subfigure}
\usepackage{url}
\usepackage{verbatim}
\usepackage{svg}
\usepackage{color}
\usepackage{booktabs}
\usepackage{makecell}
\usepackage{bbding}
\makeatletter

\newcommand{\Rmnum}[1]{\expandafter\@slowromancap\romannumeral #1@}
\makeatother

\usepackage{newfloat}
\usepackage{listings}
\def\BibTeX{{\rm B\kern-.05em{\sc i\kern-.025em b}\kern-.08em
    T\kern-.1667em\lower.7ex\hbox{E}\kern-.125emX}}
\begin{document}

\title{FedTrip: A Resource-Efficient Federated Learning Method with Triplet Regularization
}

\author{
\IEEEauthorblockN{
Xujing Li\IEEEauthorrefmark{1} \IEEEauthorrefmark{2}, 
Min Liu\IEEEauthorrefmark{1} \IEEEauthorrefmark{2}, 
Sheng Sun\IEEEauthorrefmark{1}, 
Yuwei Wang\IEEEauthorrefmark{1}, 
Hui Jiang\IEEEauthorrefmark{1} \IEEEauthorrefmark{2},
Xuefeng Jiang\IEEEauthorrefmark{1} \IEEEauthorrefmark{2}
}
\IEEEauthorblockA{
\IEEEauthorrefmark{1}Institute of Computing Technology, Chinese Academy of Sciences, Beijing, China\\
\IEEEauthorrefmark{2}University of Chinese Academy of Sciences, Beijing, China\\
% \IEEEauthorrefmark{3}Zhongguancun Laboratory, Beijing, China\\
\IEEEauthorrefmark{1}\{lixujing19b, liumin, sunsheng, ywwang, jianghui, jiangxuefeng21b\}@ict.ac.cn,
}
}

\maketitle

\begin{abstract}
In the federated learning scenario, geographically distributed clients collaboratively train a global model. Data heterogeneity among clients significantly results in inconsistent model updates, which evidently slow down model convergence.
To alleviate this issue, many methods employ regularization terms to narrow the discrepancy between client-side local models and the server-side global model. However, these methods impose limitations on the ability to explore superior local models and ignore the valuable information in historical models. 
Besides, although the up-to-date representation method simultaneously concerns the global and historical local models, it suffers from unbearable computation cost.
To accelerate convergence with low resource consumption, we innovatively propose a model regularization method named FedTrip, which is designed to restrict global-local divergence and decrease current-historical correlation for alleviating the negative effects derived from data heterogeneity.
FedTrip helps the current local model to be close to the global model while keeping away from historical local models, which contributes to guaranteeing the consistency of local updates among clients and efficiently exploring superior local models with negligible additional computation cost on attaching operations.
Empirically, we demonstrate the superiority of FedTrip via extensive evaluations. 
To achieve the target accuracy, FedTrip outperforms the state-of-the-art baselines in terms of significantly reducing the total overhead of client-server communication and local computation. 

\end{abstract}


\section{Introduction}
Over the last few decades, massive data have brought about the dramatic development of extensive Artificial Intelligence (AI) applications \cite{he2016deep, vaswani2017attention, devlin2018bert, liu2021swin}. 
In real life, data are produced by ubiquitous sensing and computing devices, such as mobile phones and wearable devices \cite{seneviratne2017survey, zhang2019deep, lim2020federated}. 
However, in the traditional centralized learning paradigm, raw data are required to be gathered from decentralized devices and transmitted to the central server, which causes unavoidable privacy disclosure and unreasonably high communication overhead.

To alleviate the above issue, Federated Learning (FL) \cite{konevcny2016federated, bonawitz2019towards, yang2019federated}, a distributed learning paradigm that enables participants to collaboratively train a global model without local data exchange, has emerged as an important paradigm and attracted a lot of research interest \cite{wang2020federated, wolfrath2022haccs, li2022federated}. 
In the fundamental FL algorithm, FedAvg \cite{mcmahan2017communication}, the clients in the FL system train the local models on their private data for multiple local iterations, and upload their updated models to the server for generating an aggregated global model.
With no data exchange and periodic model aggregation, FL has significant potential to facilitate AI applications in practice \cite{kaissis2020secure}.

Nevertheless, FL confronts a key challenge of data heterogeneity \cite{khaled2020tighter, karimireddy2020scaffold, li2020federated1}, which means that data distributions among clients follow the nonindependent and identically distributed (non-IID) characteristic. This phenomenon inevitably causes apparent update inconsistency among local models \cite{zhao2018federated, hsieh2020non} and a decline in model generalizability \cite{kairouz2019advances}. 
As a result, the resource consumption for training a model to achieve the desired performance tends to remarkably increase. 

To date, various kinds of approaches have been proposed to mitigate the impact of data heterogeneity \cite{wang2020tackling, reddi2020adaptive, li2021fedbn}. 
Among them, model regularization \cite{li2020federated, acar2021federated, li2019feddane, kim2022communication} is the general solution, which focuses on constraining the training divergence between the global model and local models by introducing regularization terms into the local loss function. 
However, constraining model update inherently limits the convergence potential in the local training process \cite{mendieta2022local}. 
The regularization terms constrain the update divergence but directly prevent the local model from exploring the parameter space far from the global model, where there possibly exists useful information that helps discover the superior local models and promotes model convergence. 
In addition, the above methods overlook the useful model information that can be learned from historical local models, which is viewed as the diversity of knowledge representation among local models. Consequently, insufficient model information utilization causes slow convergence.

To sufficiently utilize the information from historical local models, a model representation method MOON \cite{li2021model} is proposed.
It modifies model updates via designing a loss function based on contrastive learning, whose input terms are the representational outputs of the global model, the current local model, and the historical local model. 
However, it requires a mass of feedforward operations for extracting feature representation and leads to tremendous computation cost.
To date, there is no method that is able to sufficiently utilize the model information with low resource consumption, aiming to settle data heterogeneity.

Motivated by the limitations of existing studies, we propose a novel model regularization method named FedTrip. We expand the triplet loss function \cite{schroff2015facenet} to the model level for measuring the divergence of model parameters in the model regularization style.
% **************************
Specifically, a triplet regularization term is added to the local loss function. This term helps the current model to stay close to the global model for guaranteeing update consistency and keep away from historical local models for efficiently exploring parameter spaces.
% **************************
Our proposed FedTrip is able to efficiently extract helpful convergence information during the training process with attaching minor operations, negligible computation cost, and no additional communication cost. 
Compared to existing methods, FedTrip achieves sufficient model information utilization and realize convergence acceleration with very low resource consumption under data heterogeneity.

The main contributions are summarized as follows:
\begin{itemize}
    \item 
    In order to overcome the impact of data heterogeneity with low resource consumption, we propose a novel and effective model regularization method in FL under the circumstance of data heterogeneity, named FedTrip. 
    Specifically, we introduce a triplet regularization term into the local loss function. This term decreases the global-local convergence discrepancy and simultaneously increases the current-historical model difference with negligible computation cost and no additional communication cost, which can guarantee consistent model updates and obtain more useful training information.
    \item In addition, we theoretically analyze the convergence property of FedTrip based on easily-satisfied convergence conditions in the FL system. Theoretical results demonstrate that FedTrip can achieve faster convergence with given hyperparameters than FedProx.
    \item Extensive experiments are conducted to verify the performance of FedTrip. Experimental results show its superiorities in terms of client-server communication overhead and local computation overhead under various settings and hyperparameters. Especially, FedTrip is satisfactory under strict settings of FL, including highly-skewed data heterogeneity and low client participation ratio.
\end{itemize}

\section{Related Work}
\begin{figure}[!t]
% \setlength{\abovecaptionskip}{0pt}
% \subfigbottomskip=3pt
% \subfigcapskip=-4pt
  \centering
    \subfigure[IID setting]{\includegraphics[width=1.7in]{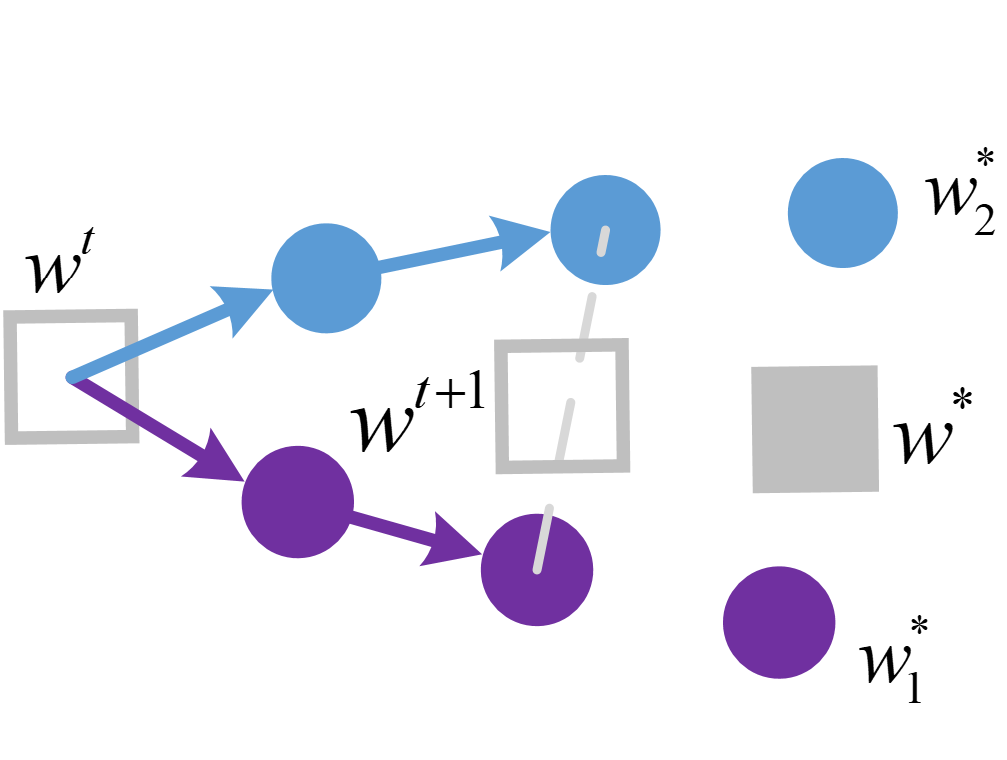}}
    \subfigure[non-IID setting]{\includegraphics[width=1.7in]{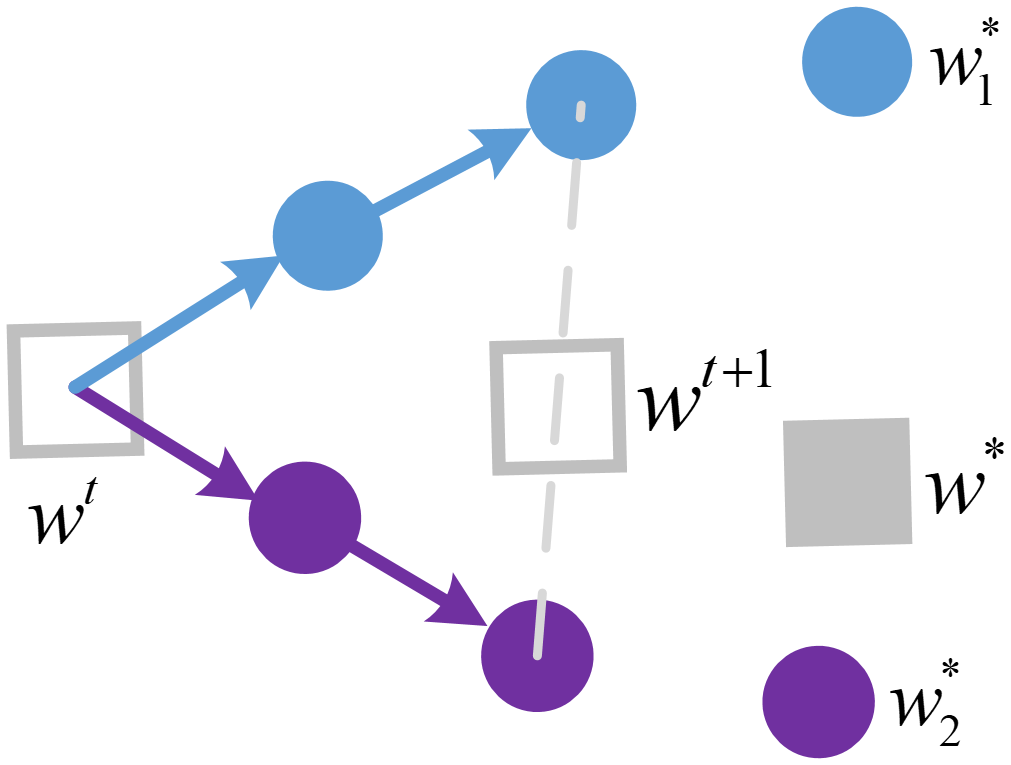}}
  \caption{Illustration of model updates in federated learning with IID data and non-IID data settings. Circles are the local model updates and optima. Rectangles are the global model updates and optimum.}
  \label{fig_Hete_illus}
  % \vspace{-1em}
\end{figure}

In the FL training paradigm, clients are randomly selected to execute local training at each communication round, during which clients' local data are not allowed to share with others. 
Following \cite{zhao2018federated}, we illustrate the impact of data heterogeneity on local model updates as shown in Fig. \ref{fig_Hete_illus}. 
When local data are IID, for each client $k$, the local optimum $w^*_k$ is close to the global optimum $w^*$, and the updates of $w^t_k$ are consistent with other clients. 
When local data are non-IID, the local optimum $w^*_k$ cannot align with the global optimum $w^*$, and the updates of $w^t_k$ have inconsistency with others. 
This inevitably causes the obvious divergence of the local models.
Although this issue has attracted many research interests, 
we only focus on the related methods that intuitively inspire us to tackle the impact of data heterogeneity. 
\subsection{Model Regularization} Recent studies have focused on utilizing model regularization to mitigate the impact of data heterogeneity in FL.
In these works, the local training objective of clients not only measures the empirical risk over local data but also attaches additional regular terms to reduce the training divergence among local models.
FedProx \cite{li2020federated} is the first study that adds a proximal term to effectively limit the local model updates by restricting local models to approximate the global model. 
Based on FedProx, FedDANE \cite{li2019feddane} further leverages a gradient correction term to improve training performance.
Although FedDANE has more regularization terms and encouraging theoretical guarantee, it consistently underperforms FedProx in evaluations. 
FedDyn \cite{acar2021federated} dynamically updates the local loss function by adding a term to guarantee the similarity between local gradients and the parameters, which ensures that the local optima are asymptotically consistent with the global optimum. 

However, existing model regularization methods mainly contribute to directly constraining the discrepancy between local models and the global model via adding regularization terms into the loss function, which potentially prevents local models from exploring the superior parameters and obtaining more useful convergence information. 
Moreover, they overlook the information in historical local model, leading to insufficient information utilization and unsatisfactory model performance. 

\subsection{Model Representation} 
A few recent studies also pay attention to tackling the issue of data heterogeneity in FL using model representation, which devote to optimizing specific loss functions based on modifying feature representations of the local models under the guidance of global model representation.
FedGKD \cite{yao2021local} aligns the feature representations of the global model and local models via knowledge distillation \cite{hinton2015distilling}, which achieves relatively consistent local and global representations by guiding local model training through the global model. 
However, this method still overlooks the information in historical local model. 
MOON \cite{li2021model} takes historical information into consideration and designs a model-contrastive loss function based on contrastive learning \cite{hadsell2006dimensionality} to tune the feature representation similarity among the global model, the current local model, and the historical local model.
However, this method may not be practical because it requires 3$\times$ feedforward computation operations for calculating feature representations.

\begin{table}
    \renewcommand{\arraystretch}{1.2}
    \centering
    \caption{Comparison of existing methods on information utilization and resource cost of method operations.}
    \begin{tabular}{ccc}
    \Xhline{2\arrayrulewidth}
    {\bf Methods} & {\bf Information utilization} & {\bf Resource cost}\\
    \hline
    Model regularization & Insufficient & {\bf Low} \\
    Model representation & {\bf Sufficient} & High \\
    FedTrip & {\bf Sufficient} & {\bf Low} \\
    \Xhline{2\arrayrulewidth}
    \end{tabular}
    \label{tab:methods comparison}
  \vspace{-0.5em}
\end{table}

Different from the above methods, we design a novel triplet regularization term inspired by the triplet loss \cite{schroff2015facenet}, providing the insight of narrowing the convergence divergence between the current local model and the global model to guarantee the consistency training update while putting the current local model away from historical local models for exploring the superior parameters.
As shown in Table \ref{tab:methods comparison}, our proposed FedTrip integrates the advantages of model regularization and model representation methods, aiming to achieve sufficient information utilization with negligible computation cost. 
The quantitive analysis of computation and communication consumption at each communication round of related methods is presented in Appendix \ref{FirstAppendix}.

\section{Problem Formulation and Theoretical Assumptions}

\subsection{Problem Formulation}
Without loss of generality, we assume a FL system consisting of $N$ clients and a central server.
Let $\mathcal{N}=\{1, 2, \cdots, N\}$ denote the set of clients, and the private data that each client $k\in\mathcal{N}$ stores are denoted by $\mathcal{D}_k$.
Considering of data heterogeneity, data distributions across clients differ and follow non-IID in our setting.
The goal of our system is to minimize the average loss over heterogeneous data sampled from distributed clients, which is expressed as:
\begin{equation} 
    \min_{w\in \mathbb{R}^d} f(w) = \min_{w\in \mathbb{R}^d} \frac{1}{N} \sum_{k = 1}^{N} F_k(w; \mathcal{D}_k),
\end{equation}
where $w$ is the parameters of the global model, and $F_k(w;\mathcal{D}_k)$ measures the local empirical risk on client $k$ over $\mathcal{D}_k$.

The whole process of FL splits into multiple communication rounds.
At the $t$-th round, the server first randomly selects a fixed number $K$ of clients, denoted as $\mathcal{S}^{t}$, to participate in training. 
After client selection, the server synchronously transmits the global model $w^{t-1}$ to the selected clients. 
Afterwards, all clients in $\mathcal{S}^{t}$ perform local model training based on their private data in parallel, and generate the updated local models $\{w_k^{t}\}$.
All updated local models are transmitted back to the server when all clients in $\mathcal{S}^t$ finish training.
The server then aggregates local models to form the updated global model as 
\begin{equation}
    w^{t} = \sum_{k=1}^{K} a_k^{t} w_k^{t},
\end{equation}
where $a_k^{t}$ indicates the weighted coefficient of client $k$, and $\sum\nolimits_{k\in\mathcal{S}^{t}} a_k^{t}=1$.
In the fundamental FL method FedAvg \cite{mcmahan2017communication}, $a_k=\frac{\vert\mathcal{D}_k\vert}{\vert\mathcal{D}_\mathcal{S}^{t}\vert}$, where $\lvert \mathcal{D}_k\rvert$ is the number of data samples in client $k$ and $\lvert \mathcal{D}_\mathcal{S}^{t}\rvert$ is the total data size of $\mathcal{S}^{t}$.

\subsection{Theoretical Assumptions}
For the convenience of theoretical analysis, we give a few standard assumptions (see e.g., \cite{li2020federated, karimireddy2020mime}), and leverage them to conduct convergence analysis in Section IV-C.
\\
{\bf Assumption 1 (\textit{L-smooth})\label{Ass_1}}. 
The stochastic gradient of loss function at each client $k$ is \textit{L smooth}, i.e.,
\begin{align}
    \lVert \nabla F_k(w_i)- \nabla F_k(w_j)\rVert < L\lVert w_i-w_j \rVert\ \forall{w_i,w_j} \in \mathbb{R}^d.
\end{align}
% {\bf Assumption 2 (Bounded Gradient)\label{Ass_2}}. 
% The norm of the stochastic gradient of loss function at each client $k$ has the upper bound $G$ for any $w\in \mathbb{R}^d$, i.e.,
% \begin{align}
% 	\lVert\nabla F_k(w) \rVert\leq G.
% \end{align}
{\bf Assumption 2 (Bounded Gradient Dissimilarity)\label{Ass_2}}. 
The norm of stochastic gradients between the loss function of each client $k$ and the global objective function are bounded, i.e.,  
\begin{align}
    \lVert\nabla F_k(x)\rVert ^2\leq B^2\lVert\nabla f(x)\rVert ^2.
\end{align}
\par

\section{Method}
In this section, we first highlight our motivation of designing FedTrip, and then elaborate on the details of FedTrip. 
Finally, we conduct the convergence analysis of FedTrip.
\begin{figure}[!t]
% \setlength{\abovecaptionskip}{-3pt}
% \subfigbottomskip=2pt
% \subfigcapskip=-4pt
  \begin{center}
    \subfigure[Server, 50 rounds]{\includegraphics[width=0.32\linewidth]{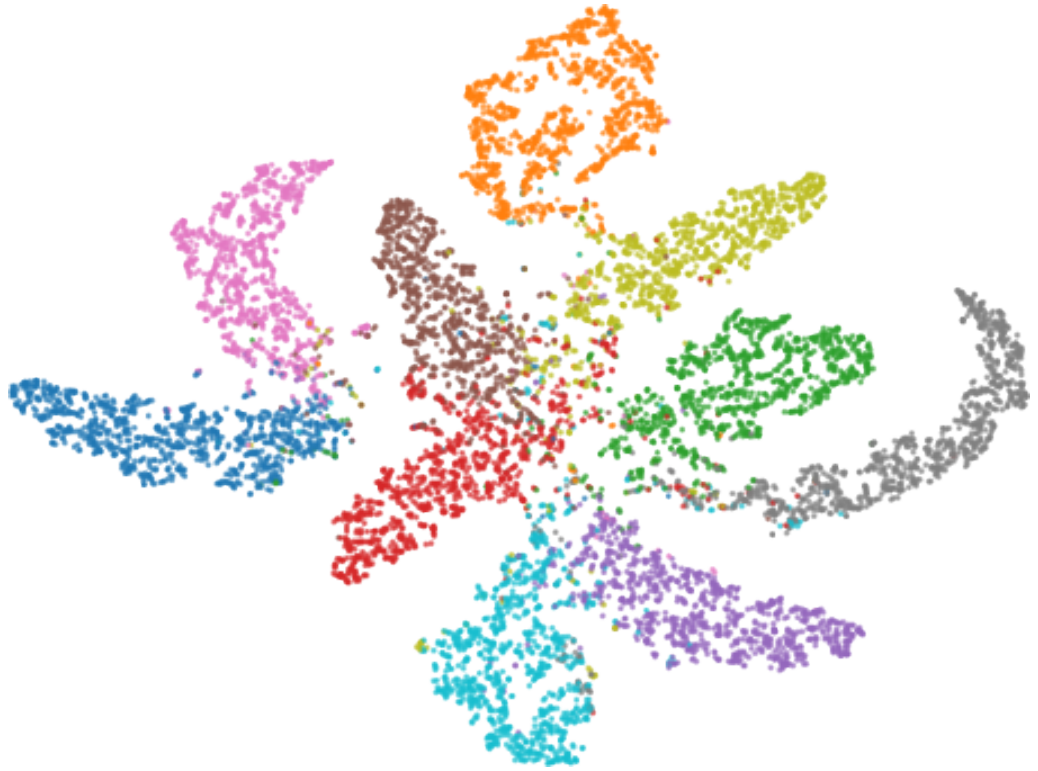}}
    \subfigure[Client 1, 50 rounds]{\includegraphics[width=0.32\linewidth]{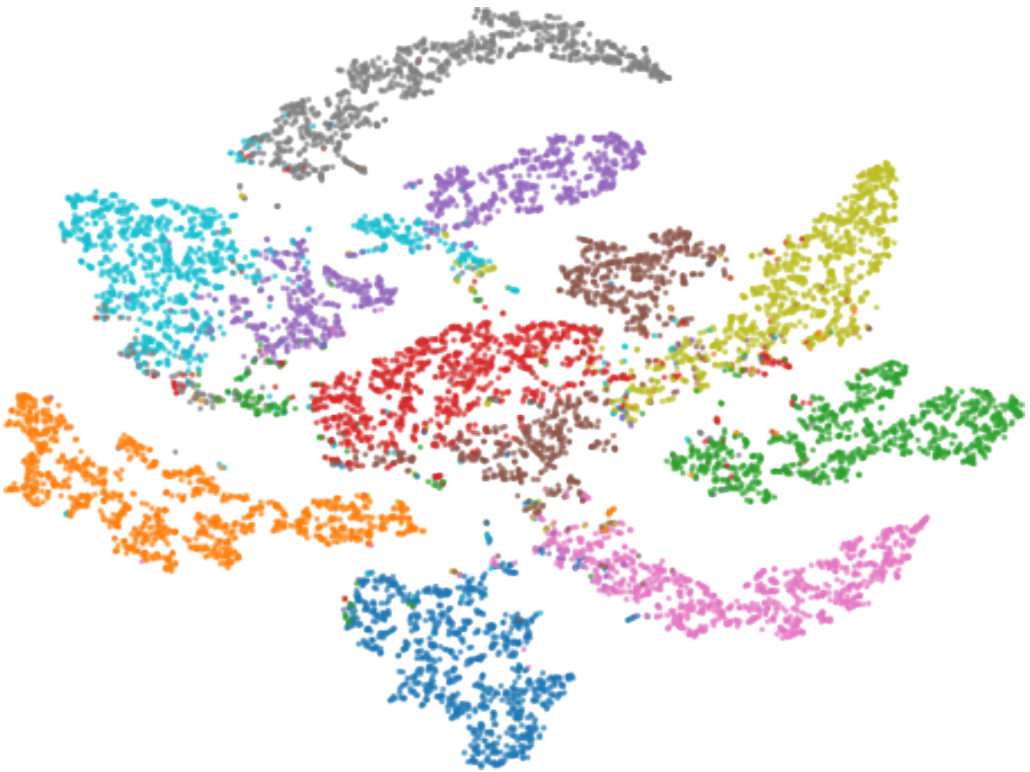}}
    \subfigure[Client 1, 30 rounds]{\includegraphics[width=0.32\linewidth]{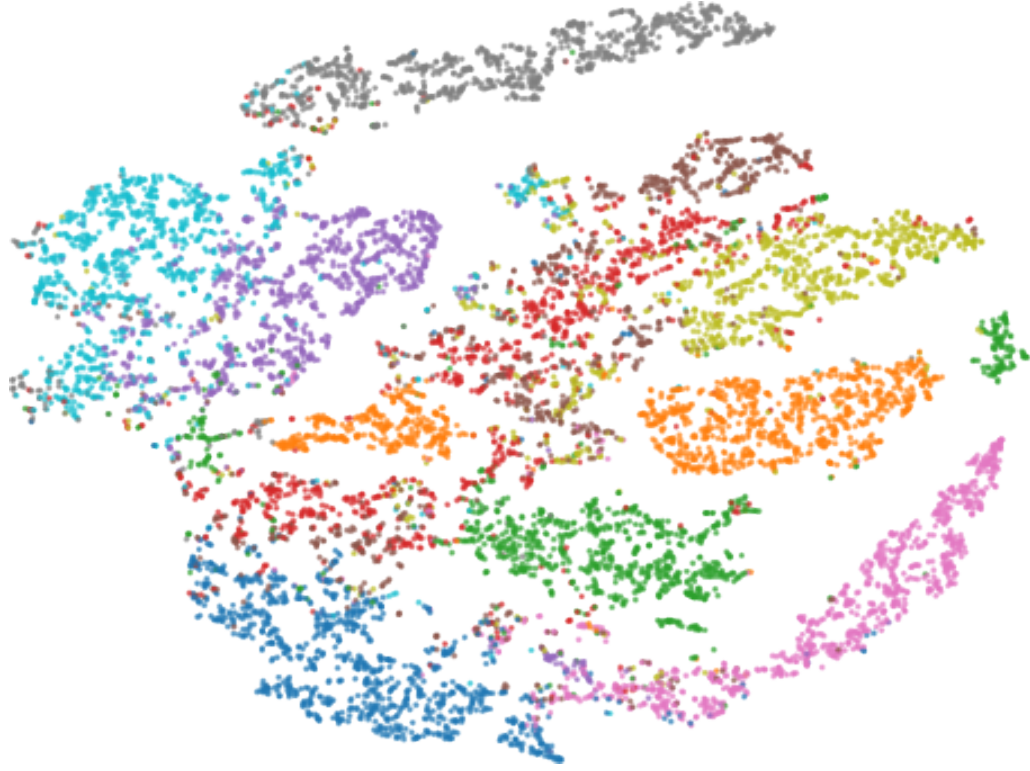}}
  \end{center}
  \caption{T-SNE visualization of the global model at round 50 and of the local model of client 1 at round 30 and 50 on the test dataset.}\label{fig_tsne}
  % \vspace{-1.5em}
\end{figure}

\subsection{Motivation}
To verify the intuition of our method, we train a CNN model using FedAvg\cite{mcmahan2017communication} on MNIST\cite{lecun1998gradient} (we use the default experiment settings, details of settings and the models can be seen in Section \Rmnum{5}). 
Fig. \ref{fig_tsne} displays the t-SNE\cite{van2008visualizing} visualization of features from the test dataset at the last communication round. The feature representations of all classes in the global model can be distinguished. 
However, the feature representations of some classes in the local models are still mixed (seen in Figs. \ref{fig_tsne}(b)). 
This indicates that the performance of the global model is better than that of the local model. 
Moreover, along the training process, the newer local model tends to outperform the older one (seen in Figs. \ref{fig_tsne}(b), \ref{fig_tsne}(c)). 
In this view, the performance of the current local model is better than that of historical local models as well. 

Depending on this observation, existing model regularization methods have better performance by closing the current local model to the global model via constraining model updates. 
Nevertheless, the information on historical local models has been overlooked so far in the existed studies. 
A recent model representation method, MOON \cite{li2021model}, takes historical model information into consideration. 
However, 
MOON is not resource-efficient because of the tremendous computation cost related to feedforward representation operation. 
Therefore, a focus on designing a method that concurrently considers local-global divergence and current-historical correlation with low computation cost should be the primary concern. 

\subsection{Method Description}
Based on the above analysis and discussion, we provide a novel insight of adding a triplet regularization term into the local loss function of clients, inspired by \cite{schroff2015facenet}. 
%%%%%%%%%%%%%%%%%
The triplet loss is originally designed in \cite{schroff2015facenet} to decrease the distance of similar samples and increase that of distinct samples, and we expand it to the model level.
%%%%%%%%%%%%%%%%%
Specifically, we inventively propose a triplet-based loss function, which simultaneously constrains the current local model to be close to the global model as well as keeping away from historical local model, with negligible computation cost on attaching operations. 
The loss function of our FedTrip is defined as:
\begin{align}\label{eq6}
    \notag \mathcal{L} =  F(w) + & \frac{\mu}{2}\left[\lVert w_{local} - w_{global}\rVert^2\right. \\
                                  & - \left. \xi\lVert w_{local} - w_{historical}\rVert^2 \right].
\end{align}
In equation \ref{eq6}, the first term $F(w)$ represents the original local loss function.
The second term $\lVert w_{local} - w_{global}\rVert^2$ intends to guide local models closer to the global model, which keeps the consistency updates among local and global models. The third term $- \lVert w_{local} - w_{historical}\rVert^2$ intends to bring current local models away from historical local models, bringing the benefits that the current local model enables to search for the superior parameters and obtains more convergence information. 
$\mu$ and $\xi$ are the hyperparameters to measure the effect of the latter two terms. 
Note that, the value of $\xi$ is set as the interval between the current round and the last round of participating in training. 

\begin{algorithm}
\caption{FedTrip} 
\label{Algorithm_FedTrip}
\hspace*{0.02in} {\bf Input:} 
the global round $T$, the learning rate $\alpha$, the coefficients $\mu$ and $\xi$ \\
\hspace*{0.02in} {\bf Output: } 
the final model $w^T$ 
\begin{algorithmic}[1]
\FOR{$t = 1$ to $T$} % For \STATE Server Randomly selects $k$ clients as $\mathcal{D}_t$ and delivers global weight $x_t$ to them
    \STATE The server randomly selects $K$ clients as $\mathcal{S}^t$ and delivers the global model $w^{t-1}$ to them
    \FOR {client in $\mathcal{S}^t$ in parallel}
        \STATE Let $w_k^t=w^{t-1}$, and load the historical local model $\tilde{w}_k^{t-1}$
        \FOR {batch data $\zeta^t_k$}
          \STATE Calculate local loss $F_k(w^t_k;\zeta^t_k) $
          \STATE $h^t_k= \nabla F_k(w^{t}_k;\zeta^t_k) + \mu \left((w^{t}_k - w^{t-1}) \right.+\left. \xi (\tilde{w}_k^{t-1} - w_k^t)\right)$
          \STATE $w_k^{t} \ =\ w_k^{t} - \alpha \mathcal{U}(h_k^{t})$
          \ENDFOR
    \ENDFOR
    %\STATE $w_k^{t+1}=w_k^t$
    \STATE Clients in $\mathcal{S}^t$ upload $w_{k}^{t}$ to server
    \STATE The server aggregates the local models via $w^{t} = \sum_{k\in \mathcal{S}^t} \rho_k w_k^{t}$
\ENDFOR
\RETURN $w^T$
\end{algorithmic}
\end{algorithm}

The details are summarized in Algorithm \ref{Algorithm_FedTrip}.
At the beginning of the $t$-th communication round, the server randomly selects $K$ clients named $\mathcal{S}^t$ and delivers the global model $w^{t-1}$ to these clients. We denote $w_k^t$ as the current local model and $\tilde{w}_k^{t-1}$ as the historical model at client $k$, which is generated at the last local training.
After receiving $w^{t-1}$, each selected client begins its local training (line 5). 
In line 6, client $k$ trains its local model with mini-batch $\zeta_k^t$, calculates the original loss $F_k(w_k^t)$. Then local gradients are generated by the original loss value and regularization items (line 7).
After obtaining the gradients of clients, each selected client updates its local model according to the specific optimization algorithm $\mathcal{U}$ (line 8). 
The clients in $\mathcal{S}^t$ upload local models to the server when all clients finish local training.
Finally, the server aggregates the uploaded local models to obtain the updated global model $w^t$ (line 12).

%  设想了这样一个case，换成FedAvg，FedProx，FedTrip三者的比较，拼成一个大图。

\begin{figure}[t]
% \setlength{\abovecaptionskip}{-3pt}
% \subfigbottomskip=2pt
% \subfigcapskip=-6pt
  \centering
  \includegraphics[width=\linewidth]{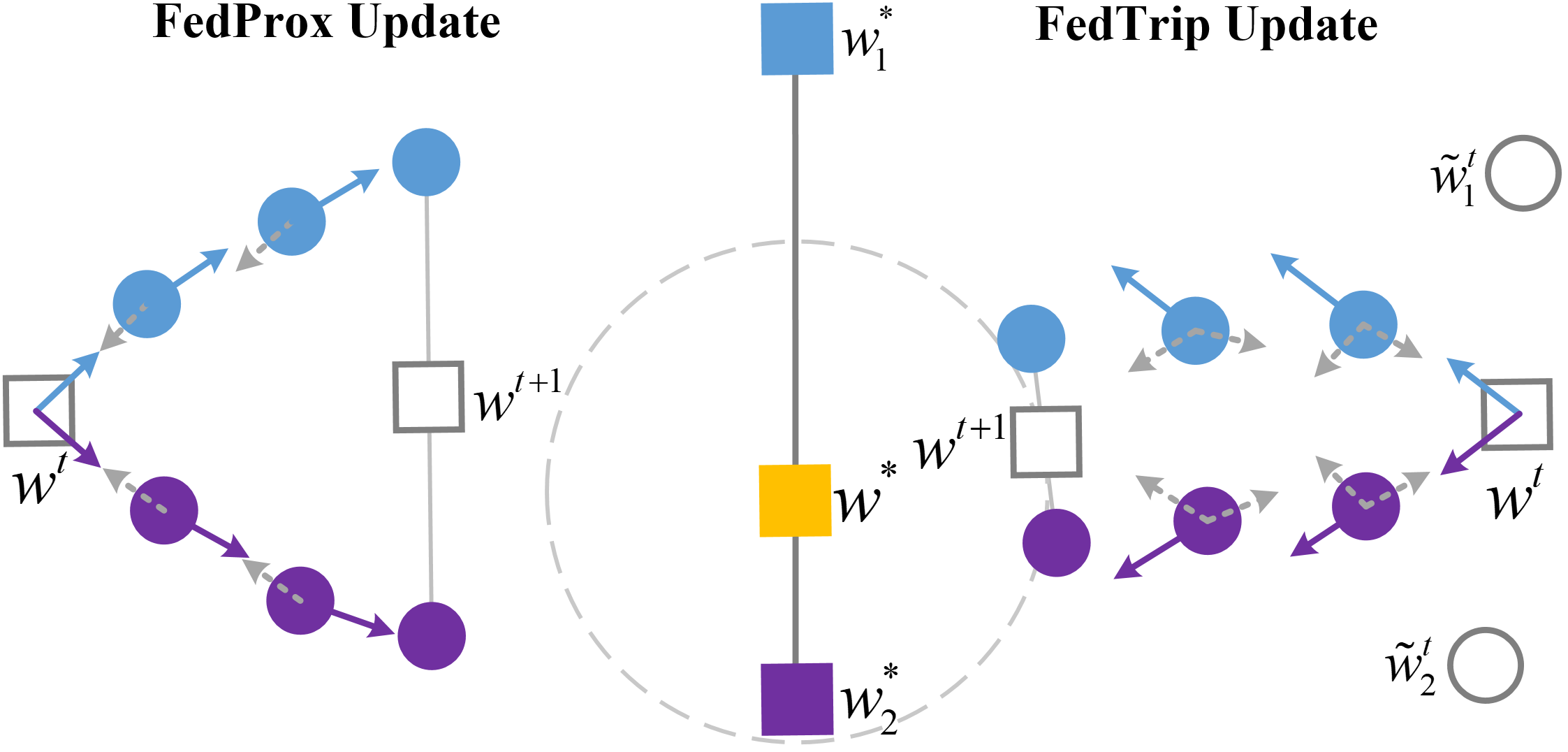}
  \caption{Illustration of model updates of 2 clients with 3 steps in FedProx and FedTrip.}
  \label{fig_alg_illus}
  % \vspace{-1.5em}
\end{figure}
Fig. \ref{fig_alg_illus} intuitively depicts the benefits of FedTrip. 
$w_1^*$ and $w_2^*$ represent the empirical local optima of client 1 and client 2 respectively, and $w^*$ represents the global optimum. 
The local SGD updates move towards the average of clients' local optimum $\frac{w_1^*+w_2^*}{2}$, which is obviously different from $w^*$. 
In the typical model regularization method, FedProx, the local SGD updates of client $k \in \{1, 2\}$ are constrained by $w^t_k-w^t$. 
However, this method potentially limits the convergence process, as the projection of local gradients towards the direction of the local optimum can be partially counteracted by the gradients generated by the regularization term. 
Creatively, in FedTrip, the local gradients under the guidance of the historical model overcome the drawback of FedProx, and the local model has the potential to explore superior parameters with the guarantee of update consistency. 
Therefore, our proposed FedTrip has the advantage of absorbing more useful information, which enables the local model to move towards the global optimum $w^*$ quickly, thus accelerating training convergence.

\par

\subsection{Convergence Analysis}
We theoretically analyze the global model convergence of FedTrip. 
This analysis mainly refers to FedProx\cite{li2020federated} and FedDANE\cite{li2019feddane}.
Firstly, we define a parameter $\gamma$ to formulate the inexactness of local optimization. \\
{\bf Definition 1 ( $\gamma$-inexact optimization)}. 
Let $w^{t+1}_k$ denote the updated local model of client $k$ based on local optimization at the $t+1$-th round, and it satisfies $\lVert \nabla h(w^{t+1}_k; w^t) \rVert < \gamma\lVert \nabla F_k(w^t)\rVert$ with $\gamma\in[0, 1)$, where $ \nabla h(w^{t+1}_k;w^t)=\nabla F_k(w^{t+1}_k)+\mu\left( w^{t+1}_k - w^t ) - \xi( w^{t+1}_k - \tilde{w}_k^t )\right)$.
\\
{\bf Theorem 1} \label{Theorem_1}. Assume that the functions $F_k$ are convex, $h_k$ is $\mu$-strongly convex. 
Given by assumptions mentioned in Section III-B, we have the expected decrease in the global objective function as:
\begin{align}
	& \mathbb{E}_{S_t}[f(w^{t+1})] \leq f(w^t) - \rho \lVert \nabla f(w^t) \rVert^2 - Q^t, \\
	\notag & \rho = \left(\frac{1-\gamma B}{\mu} - \frac{L(1+\gamma)B}{\mu^2} -\frac{L(1+\gamma)^2B^2}{2\mu^2}\right).
\end{align}
where $Q^t$ is the extra items generated by the historical information item. The expectation of coefficient of $Q^t$ is proportional to the participation ratio of a client at each round $p$. Let $\gamma=0$, which means $F_k(w)$ has the exact answer, then
\begin{align}
	\notag \rho = & \frac{1}{\mu} - \frac{LB}{\mu^2} - \frac{LB^2}{2\mu^2}. 
\end{align}
If $\mu, L, \gamma, \xi$ satisfy, we have $\rho > 0$ and $Q^t>0$, and the local objective function has the expected decrease as:
\begin{align}
	\mathbb{E}_{S_t}[f(w^{t+1})] \leq f(w^{t}) - \rho \lVert \nabla f(w^t) \rVert^2 - Q^t.
\end{align}
Note that the value of $\rho$ in FedTrip is equal to $\rho$ in Fedprox, we can get the identical decrease proportional to $\lVert \nabla f(w^t) \rVert^2$ with FedProx. Besides, $Q^t$ makes $f(w)$ have the faster convergence rate of FedTrip than that of FedProx with the help of historical model information. The main coefficient in $Q^t$ is $E_k[\xi_k^t]=\frac{p\ln p}{p-1}, t\rightarrow +\infty$, where $p$ is the client participation rate. As $E_k[\xi_k^t]$ is monotonically increasing, a low $p$ demonstrates a slow convergence rate. 
The detailed convergence analysis is referred to the Appendix \ref{SecondAppendix}. 
% {\bf Corollary 2} (Convergence rate)\label{Coro_1}. 
% Let $\gamma=0$, all local problems are sovled exactly, if we set the value of $\mu\approx(CLB^2)$, $\rho\approx$.
In summary, FedTrip is a resource-efficient and fast convergence method with
given hyperparameters.

\section{Experiments}
\subsection{Experimental Settings}
We investigate our method and comparable baselines on an open-source federated learning framework Plato\footnote{\url{https://github.com/TL-System/plato}} with PyTorch\cite{paszke2019pytorch} 1.9.1 backend, whose data partition method is based on LEAF \cite{caldas2018leaf}. Our experiments are executed on a workstation with an Intel Xeon Gold 5218 CPU @ 2.30GHz, a RAM of 376 GB, and one Nvidia GeForce 3090 GPU.\\
{\bf Datasets:} We employ MNIST \cite{lecun1998gradient}, FashionMNIST (FMNIST) \cite{xiao2017fashion}, EMNIST \cite{cohen2017emnist}, and CIFAR-10 \cite{krizhevsky2010cifar} for image classification task. The datasets cover different attributes, dimensions, and numbers of categories, which are listed in Table \ref{table_dataset}. 
Thereinto, 1 channel and 3 channels indicate grayscale and RGB images, respectively. Client samples indicate the number of data samples at each client.\\
{\bf Models:} We train a MultiLayer Perceptron (MLP) on MNIST and FMNIST datasets. 
MLP consists of 2 fully connected layers with 100 and 10 neurons. The first fully connected layer is followed by ReLU activation \cite{nair2010rectified}.
A simple Convolution Neural Network (CNN) is used for training on MNIST, FMNIST, and EMNIST.
The CNN is modified based on LeNet5 \cite{lecun1998gradient}, consisting of 3 convolutional layers with 5$\times$5 filters followed by two fully connected layers with 84 and 10 neurons. 
Moreover, AlexNet\cite{krizhevsky2012imagenet} is trained on CIFAR-10.\\
{\bf Data Partitioning:} We adopt two popular non-IID data partitioning ways: Dirichlet distribution and orthogonal distribution. 
We generate two types of data heterogeneity via the following Dirichlet distribution to sample data labels.
First, each client draws a probability vector using the Dirichlet distribution with a concentration parameter $\alpha$, which corresponds to the prior data distribution of each class. 
The probability vectors are generated based on different random seeds and are used to sample data without replacement for clients. 
The sampling process does not stop until the number of data samples is assigned to the preset partition number. 
In our experiments, we implement 2 types of Dirichlet distributions with $\alpha = 0.1, 0.5$, named $Dir-\textit{0.1}$ and $Dir-\textit{0.5}$.

\begin{table}[!t]
\renewcommand{\arraystretch}{1.2}
\caption{Description of Datasets in the experiment.}
\label{table_dataset}
\centering
\begin{tabular}{ccccc}
\Xhline{2\arrayrulewidth}
\multirow{2}{*}{\textbf{Dataset}} 
 & \textbf{Total} & \multirow{2}{*}{\textbf{Classes}} & \multirow{2}{*}{\textbf{Channels}} & \textbf{Client}\\
 & \textbf{Samples} & & & \textbf{Samples}\\
\hline
 MNIST & 60,000 & 10 & 1 & 600  \\
 FMNIST & 60,000 & 10 & 1 & 1,000  \\
 EMNIST & 112,800 & 47 & 1 & 3,000  \\
 CIFAR-10 & 50,000 & 10 & 3 & 2,000  \\
\Xhline{2\arrayrulewidth}
\end{tabular}
\label{tab:dataset}
% \vspace{-1em}
\end{table}
\begin{table}[t]
\renewcommand{\arraystretch}{1.2}
    \setlength{\tabcolsep}{3mm}
    \centering
    \caption{Communication and computation statistics of models}
    \begin{tabular}{cccc}
    \Xhline{2\arrayrulewidth}
        \textbf{Model} & \textbf{Communication(MB)} & \textbf{Params(M)} & \textbf{MFLOPs}\\
    \hline
        MLP & 0.3 & 0.8 & 0.08\\
        CNN & 0.24 & 0.62 & 0.42\\
        AlexNet & 10.42 & 2.72 & 145.93 \\
    \Xhline{2\arrayrulewidth}
    \end{tabular}
    \label{table:model_description}
    % \vspace{-1.5em}
\end{table}

\begin{figure}[!t]
% \setlength{\abovecaptionskip}{-3pt}
% \subfigbottomskip=2pt
% \subfigcapskip=-6pt
\centering
\includegraphics[width=1\linewidth]{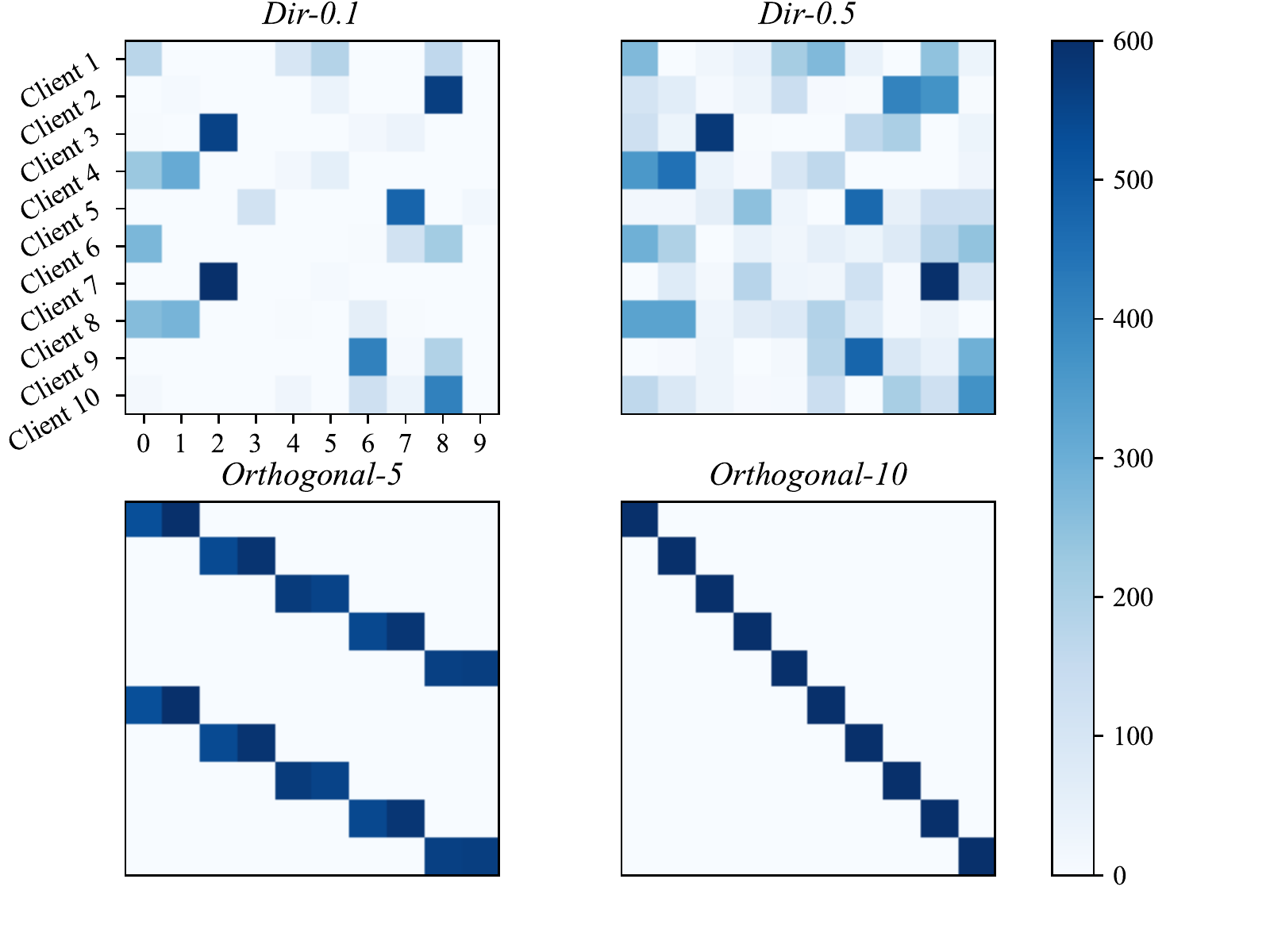}
\caption{The label distributions at clients in four settings of data heterogeneity.}
\label{fig_datapoints}
% \vspace{-1.5em}
\end{figure}

\begin{table*}[thbp]
\renewcommand{\arraystretch}{1.2}
\setlength{\tabcolsep}{0.8mm}
\caption{Comparison of communication rounds until the global model achieves the target accuracy. }
\label{table_rate_10}
\centering
\begin{tabular}{cccccccccccccccccccccc}
\Xhline{2\arrayrulewidth}
\multirow{2}{*}{{\bf Methods}} 
& \multicolumn{6}{l}{{\bf MLP}} & \multicolumn{9}{l}{{\bf CNN}} & \multicolumn{3}{l}{{\bf AlexNet}} \\
& \multicolumn{3}{l}{MNIST-87\%} & \multicolumn{3}{l}{FMNIST-75\%}
& \multicolumn{3}{l}{MNIST-90\%} & \multicolumn{3}{l}{FMNIST-75\%} 
& \multicolumn{3}{l}{EMNIST-62\%} & \multicolumn{3}{l}{CIFAR-50\%}\\
\hline
FedTrip 							& {\bf 28} & \begin{minipage}[b]
                                    {0.14\columnwidth}\raisebox{-0.1\height}{
                                    \includegraphics[width=\linewidth]{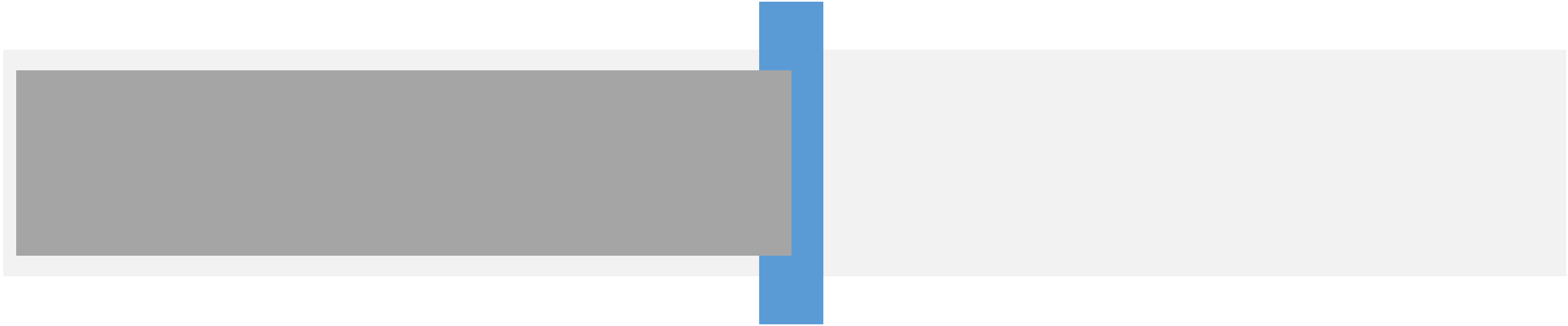}}\end{minipage} &
                                    & {\bf 9}  & \begin{minipage}[b]
                                    {0.14\columnwidth}\raisebox{-0.1\height}{
                                    \includegraphics[width=\linewidth]{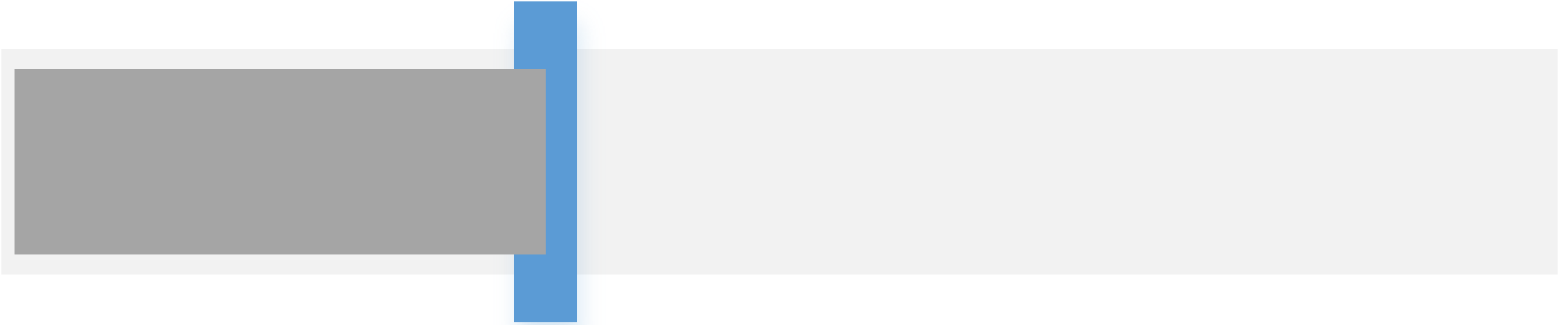}}\end{minipage}
                                    &
                                    & {\bf 24} & \begin{minipage}[b]
                                    {0.14\columnwidth}\raisebox{-0.1\height}{
                                    \includegraphics[width=\linewidth]{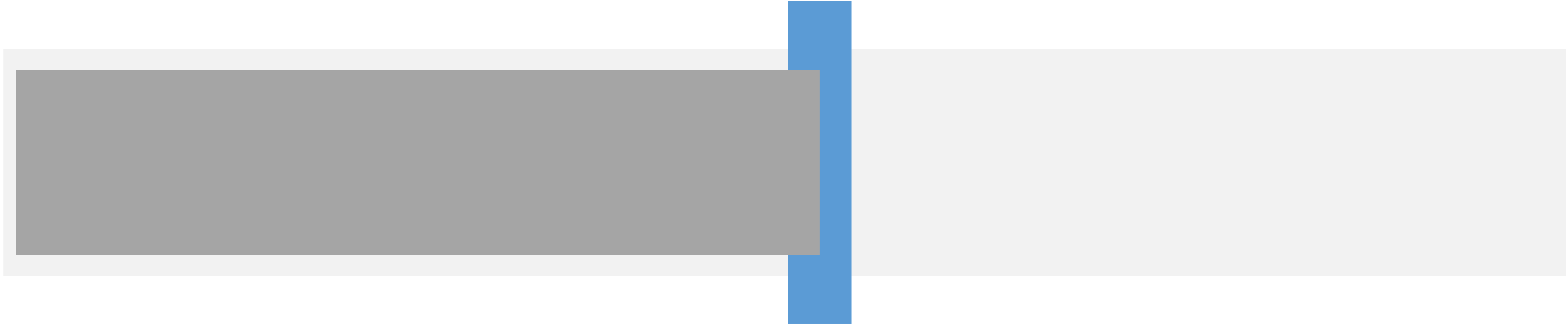}}\end{minipage} & 
                                    & {\bf 19} & \begin{minipage}[b]
                                    {0.14\columnwidth}\raisebox{-0.1\height}{
                                    \includegraphics[width=\linewidth]{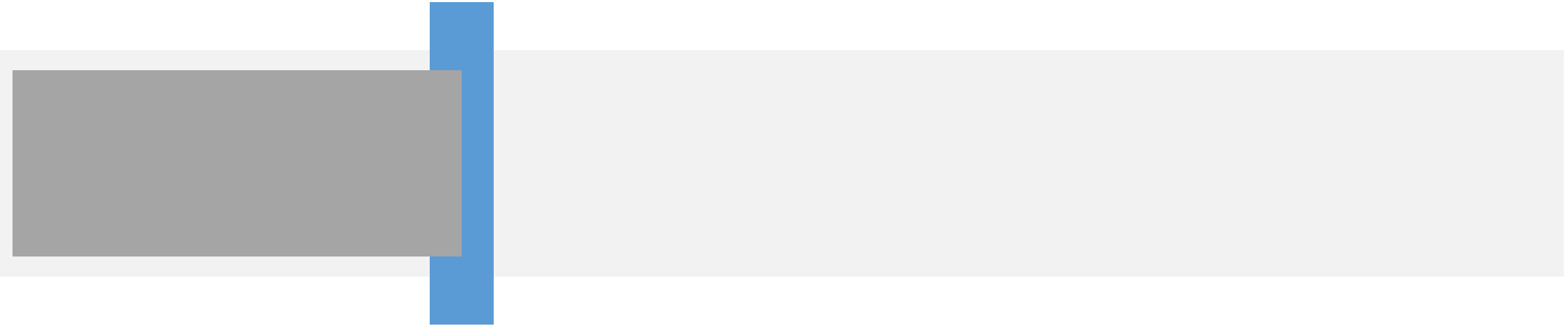}}\end{minipage}
                                    &
                                    & {\bf 32} & \begin{minipage}[b]
                                    {0.14\columnwidth}\raisebox{-0.1\height}{
                                    \includegraphics[width=\linewidth]{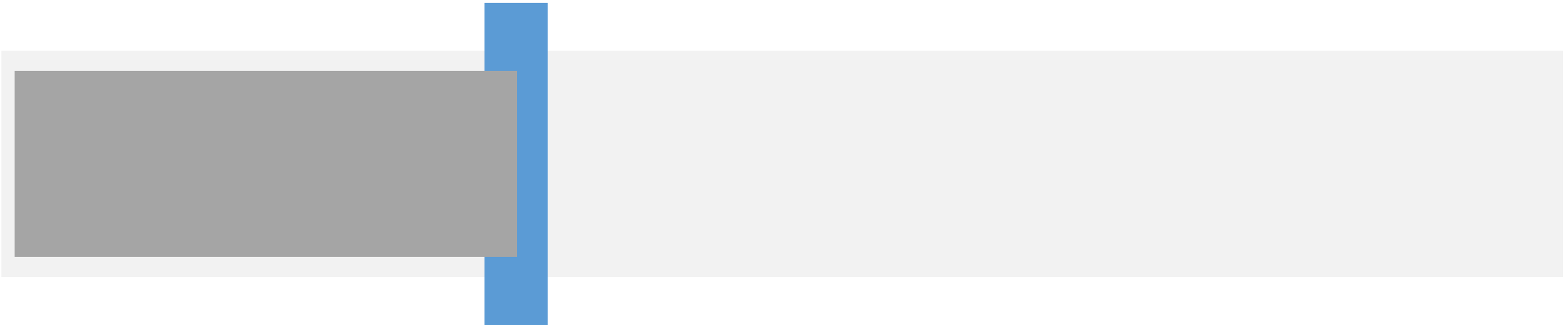}}\end{minipage} &
                                    & {\bf 46} & \begin{minipage}[b]
                                    {0.14\columnwidth}\raisebox{-0.1\height}{
                                    \includegraphics[width=\linewidth]{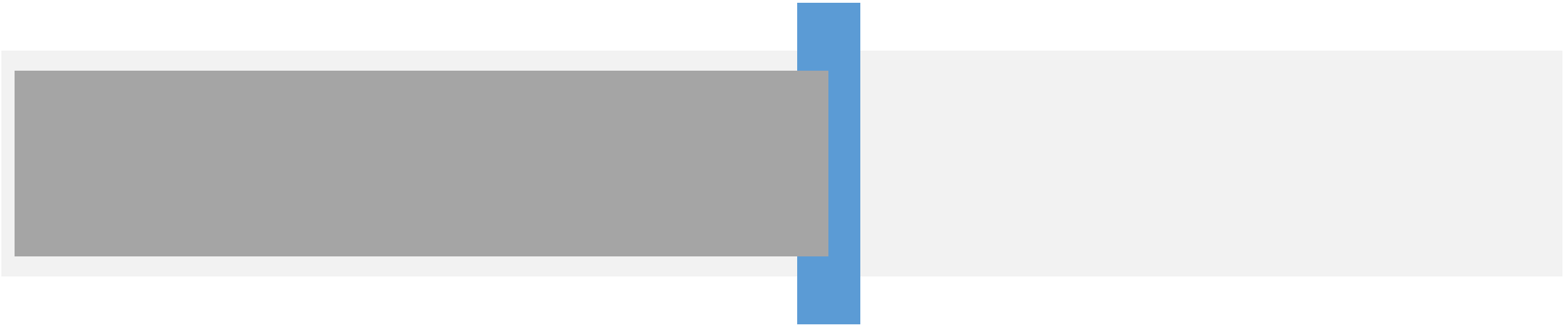}}\end{minipage} &\\
FedAvg    							& 49 & \begin{minipage}[b]
                                    {0.14\columnwidth}\raisebox{-0.1\height}{
                                    \includegraphics[width=\linewidth]{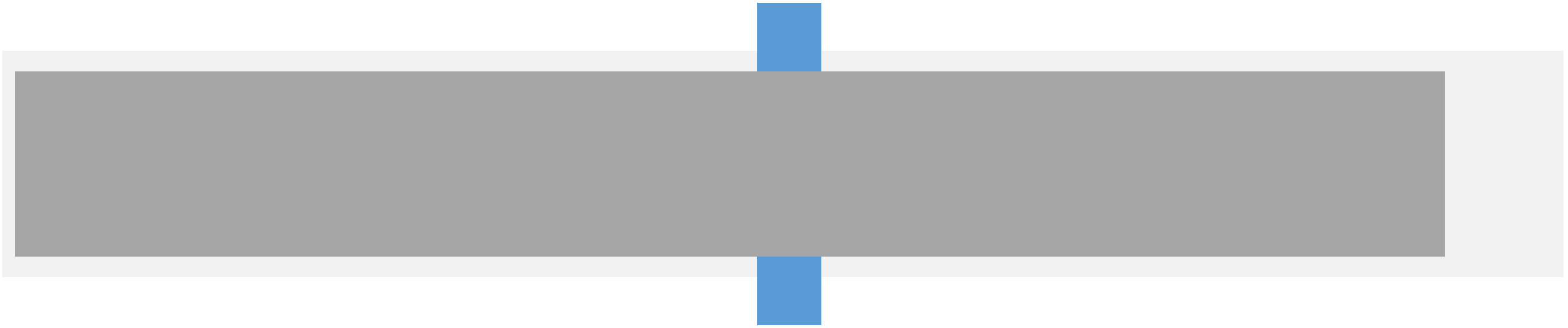}}\end{minipage}
                                    & {\color{gray} 1.75$\times$ }
                                    & 19 & \begin{minipage}[b]
                                    {0.14\columnwidth}\raisebox{-0.1\height}{
                                    \includegraphics[width=\linewidth]{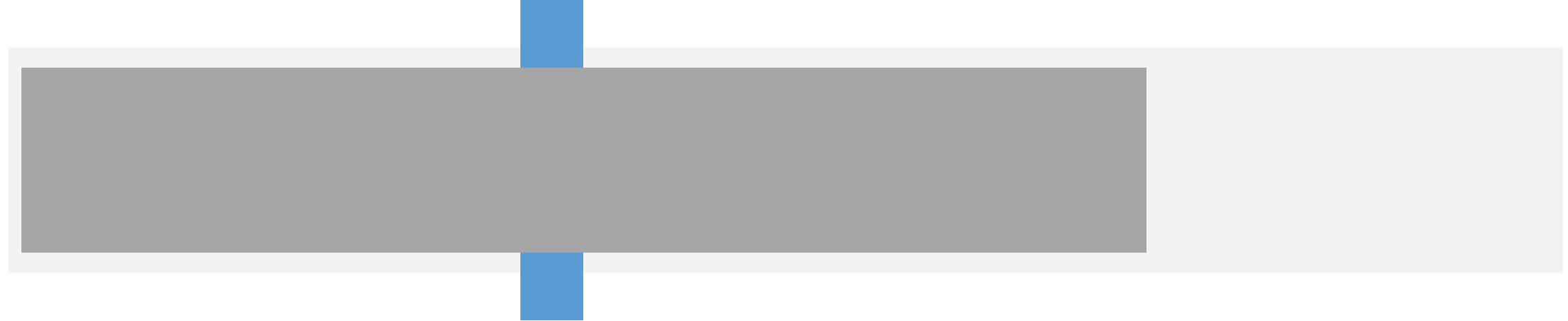}}\end{minipage}
                                    & {\color{gray} 2.11$\times$ }
                                    & 39 & \begin{minipage}[b]
                                    {0.14\columnwidth}\raisebox{-0.1\height}{
                                    \includegraphics[width=\linewidth]{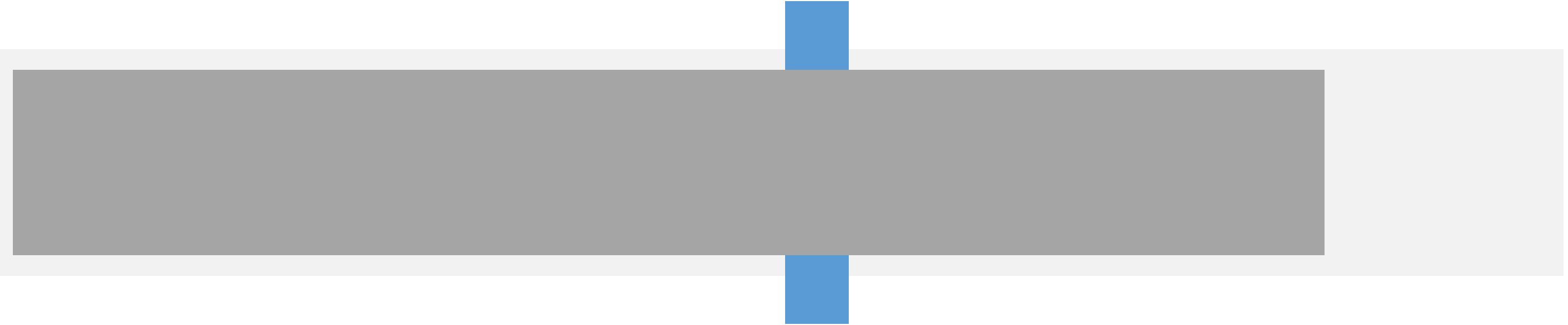}}\end{minipage}
                                    & {\color{gray} 1.63$\times$ }
                                    & 52 & \begin{minipage}[b]
                                    {0.14\columnwidth}\raisebox{-0.1\height}{
                                    \includegraphics[width=\linewidth]{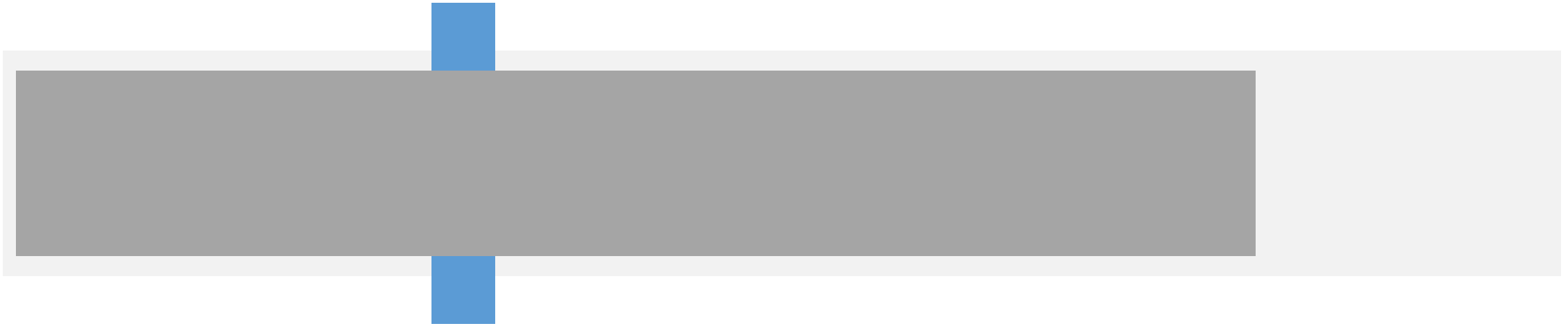}}\end{minipage}
                                    & {\color{gray} 2.73$\times$ }
                                    & 45 & \begin{minipage}[b]
                                    {0.14\columnwidth}\raisebox{-0.1\height}{
                                    \includegraphics[width=\linewidth]{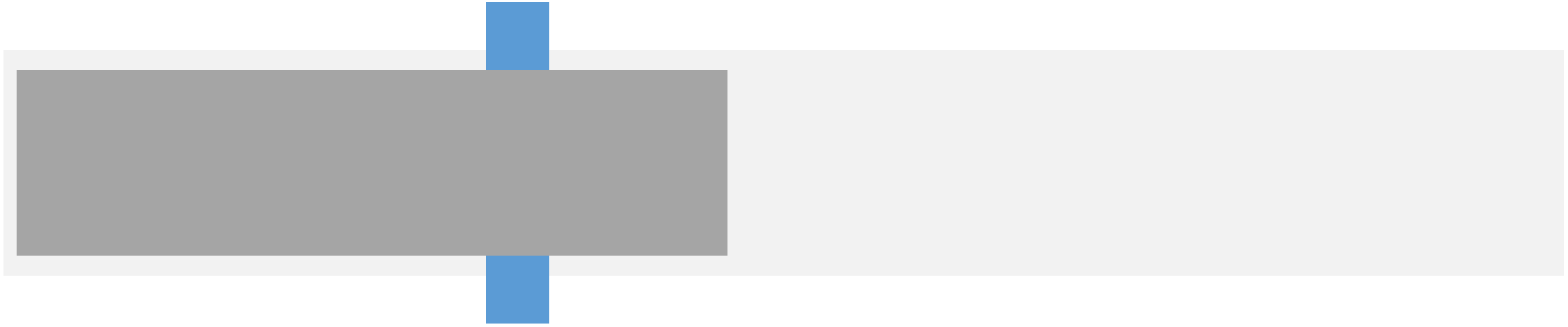}}\end{minipage}
                                    & {\color{gray} 1.4$\times$ }
                                    & 74 & \begin{minipage}[b]
                                    {0.14\columnwidth}\raisebox{-0.1\height}{
                                    \includegraphics[width=\linewidth]{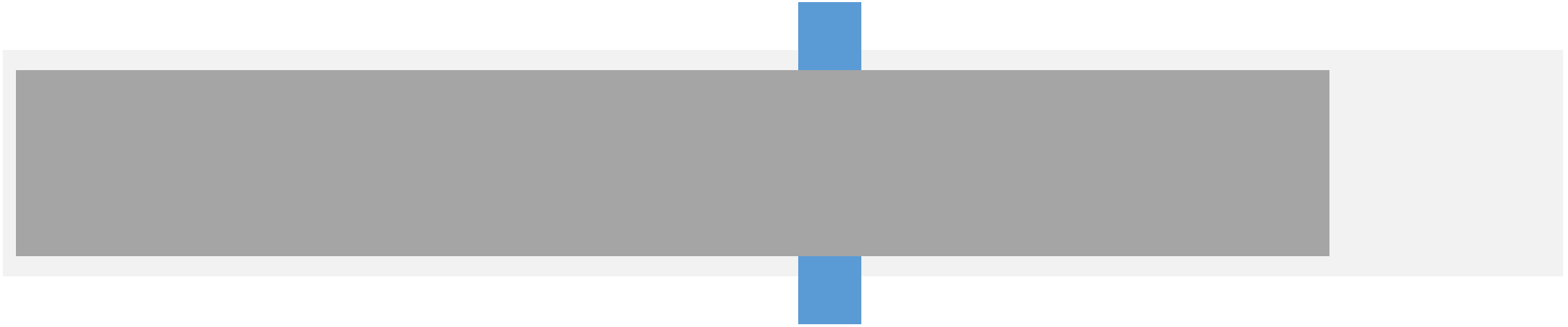}}\end{minipage}
                                    & {\color{gray} 1.61$\times$ }\\
FedProx   							& 53 & \begin{minipage}[b]
                                    {0.14\columnwidth}\raisebox{-0.1\height}{
                                    \includegraphics[width=\linewidth]{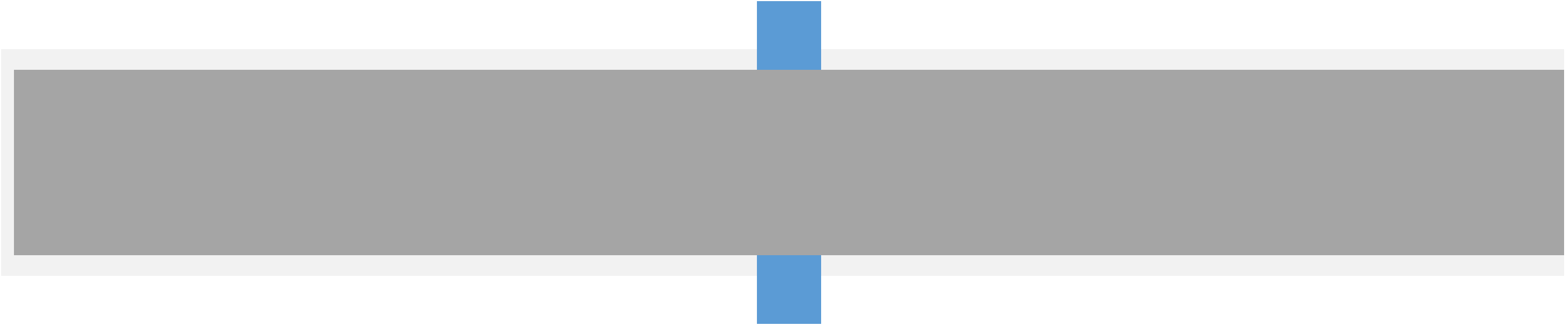}}\end{minipage}
                                    & {\color{gray} 1.89$\times$ }
                                    & 16 & \begin{minipage}[b]
                                    {0.14\columnwidth}\raisebox{-0.1\height}{
                                    \includegraphics[width=\linewidth]{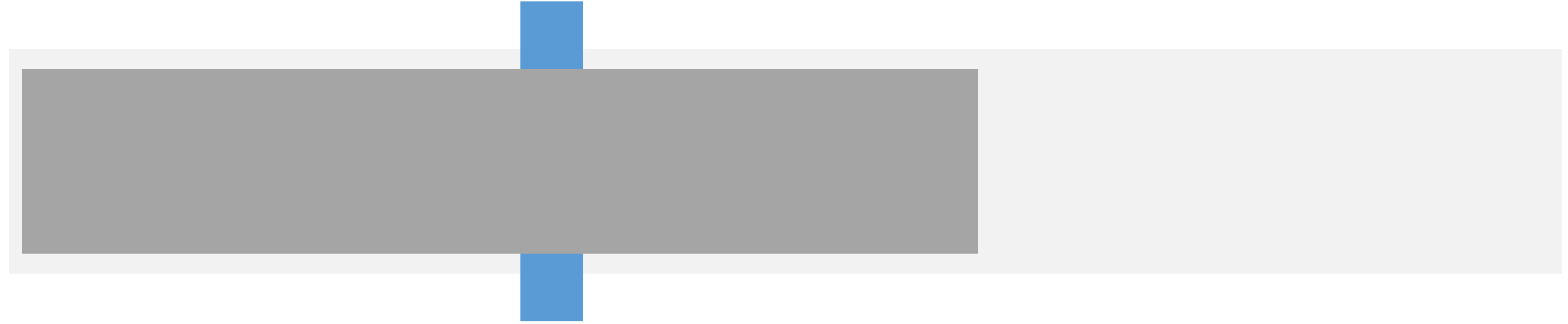}}\end{minipage}
                                    & {\color{gray} 1.78$\times$ }
                                    & 41 & \begin{minipage}[b]
                                    {0.14\columnwidth}\raisebox{-0.1\height}{
                                    \includegraphics[width=\linewidth]{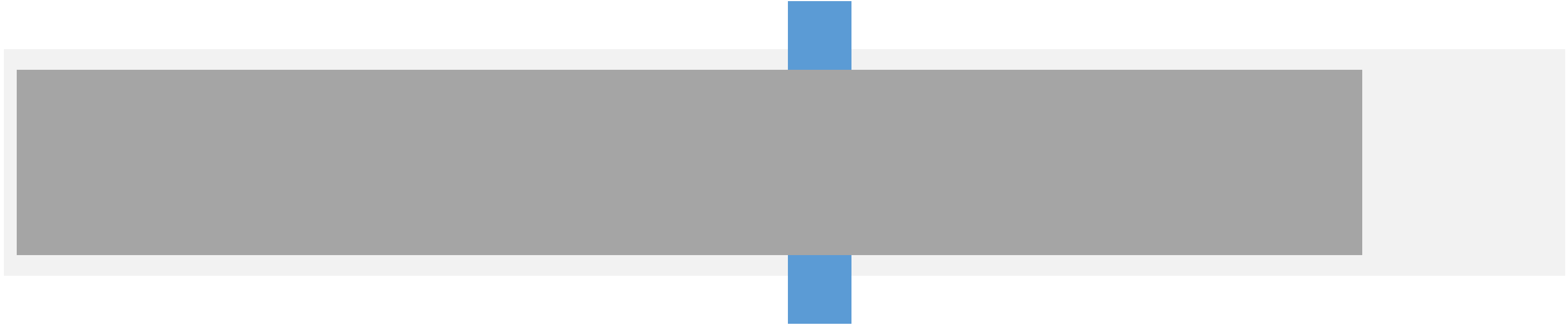}}\end{minipage}
                                    & {\color{gray} 1.71$\times$ }
                                    & 45 & \begin{minipage}[b]
                                    {0.14\columnwidth}\raisebox{-0.1\height}{
                                    \includegraphics[width=\linewidth]{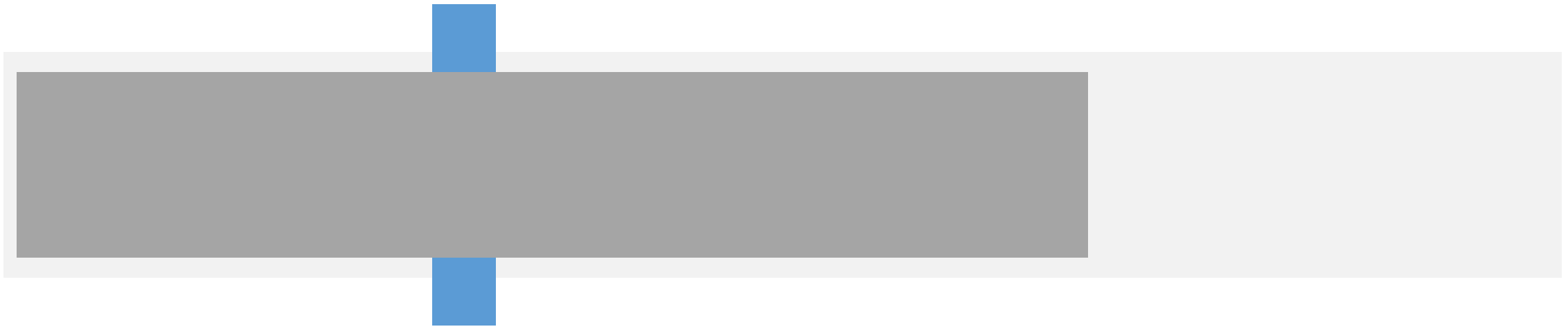}}\end{minipage}
                                    & {\color{gray} 2.37$\times$ }
                                    & 45 & \begin{minipage}[b]
                                    {0.14\columnwidth}\raisebox{-0.1\height}{
                                    \includegraphics[width=\linewidth]{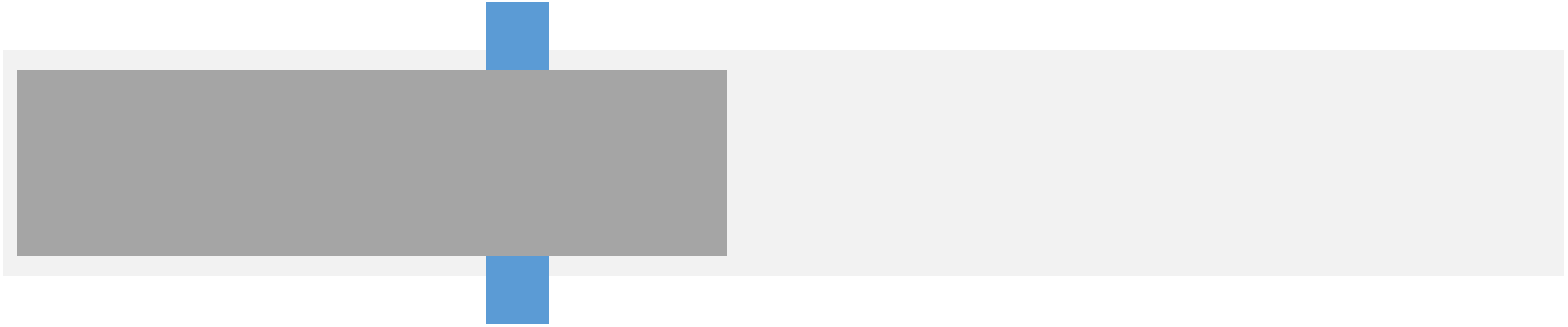}}\end{minipage}
                                    & {\color{gray} 1.4$\times$ } 
                                    & 75 & \begin{minipage}[b]
                                    {0.14\columnwidth}\raisebox{-0.1\height}{
                                    \includegraphics[width=\linewidth]{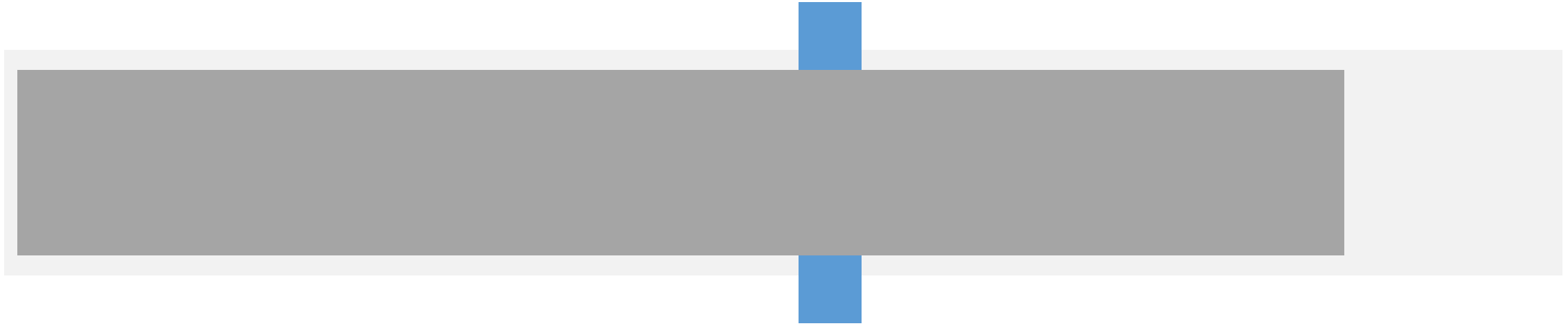}}\end{minipage}
                                    & {\color{gray} 1.63$\times$ }\\
SlowMo   					& 46 & \begin{minipage}[b]
                                    {0.14\columnwidth}\raisebox{-0.1\height}{
                                    \includegraphics[width=\linewidth]{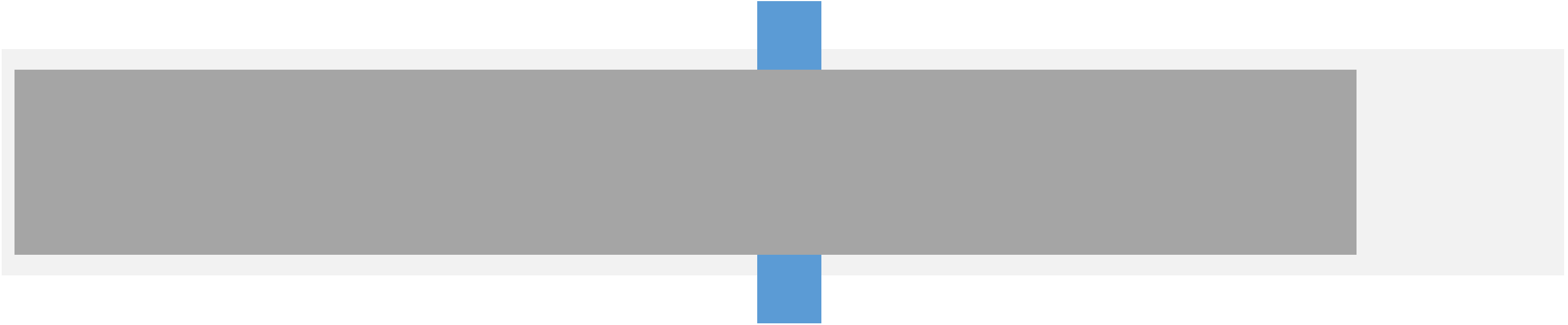}}\end{minipage}
                                    & {\color{gray} 1.64$\times$ }
                                    & 26 & \begin{minipage}[b]
                                    {0.14\columnwidth}\raisebox{-0.1\height}{
                                    \includegraphics[width=\linewidth]{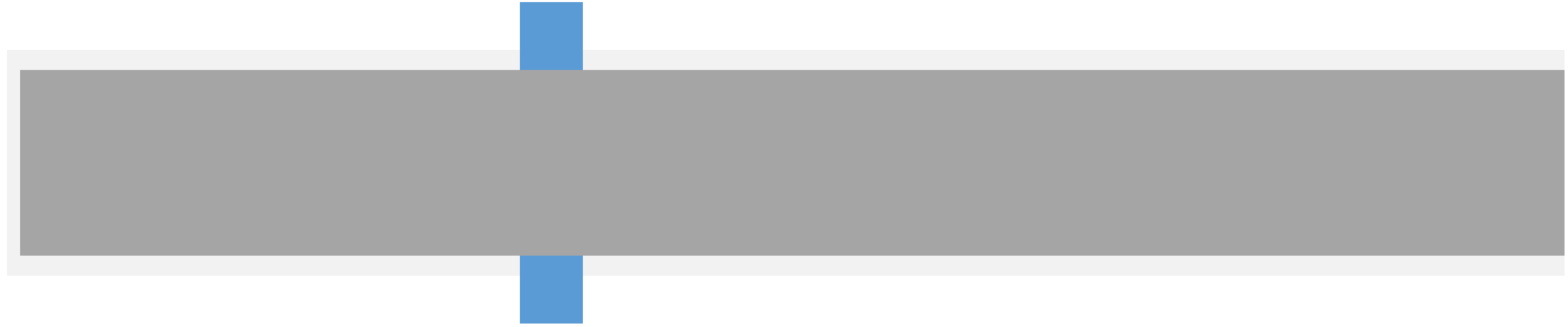}}\end{minipage}
                                    & {\color{gray} 2.89$\times$ }
                                    & 40 & \begin{minipage}[b]
                                    {0.14\columnwidth}\raisebox{-0.1\height}{
                                    \includegraphics[width=\linewidth]{MNIST_CNN_SlowMo.jpg}}\end{minipage}
                                    & {\color{gray} 1.67$\times$ }
                                    & 65 & \begin{minipage}[b]
                                    {0.14\columnwidth}\raisebox{-0.1\height}{
                                    \includegraphics[width=\linewidth]{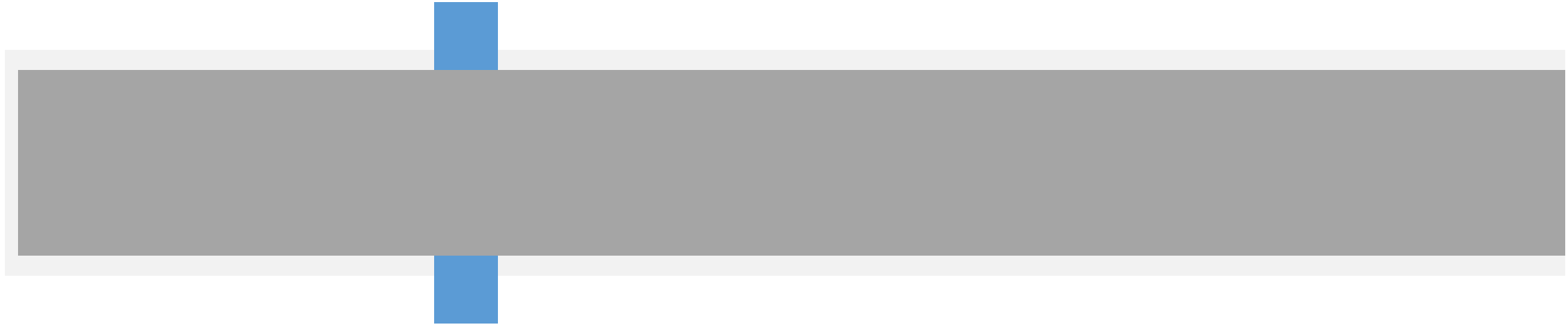}}\end{minipage}
                                    & {\color{gray} 3.42$\times$ }
                                    & 92 & \begin{minipage}[b]
                                    {0.14\columnwidth}\raisebox{-0.1\height}{
                                    \includegraphics[width=\linewidth]{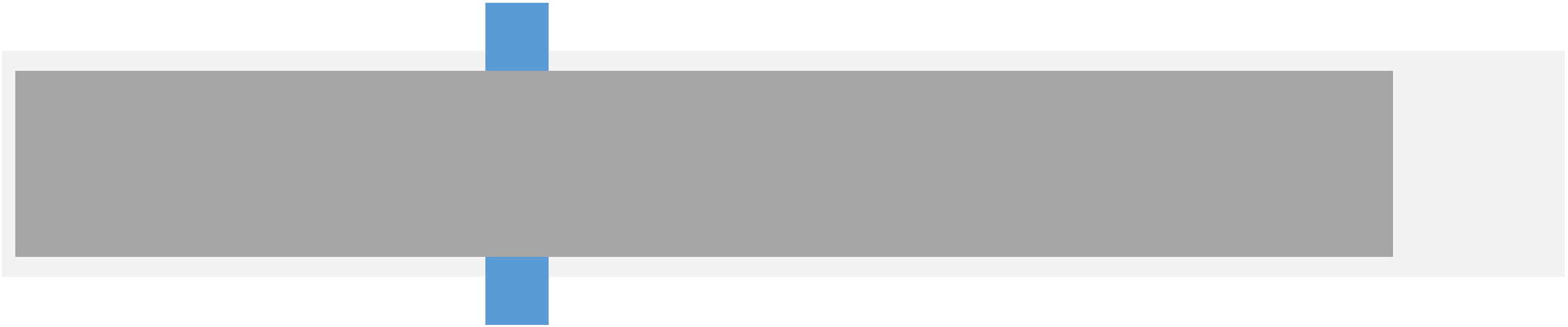}}\end{minipage}
                                    & {\color{gray} 2.88$\times$ }
                                    & 87 & \begin{minipage}[b]
                                    {0.14\columnwidth}\raisebox{-0.1\height}{
                                    \includegraphics[width=\linewidth]{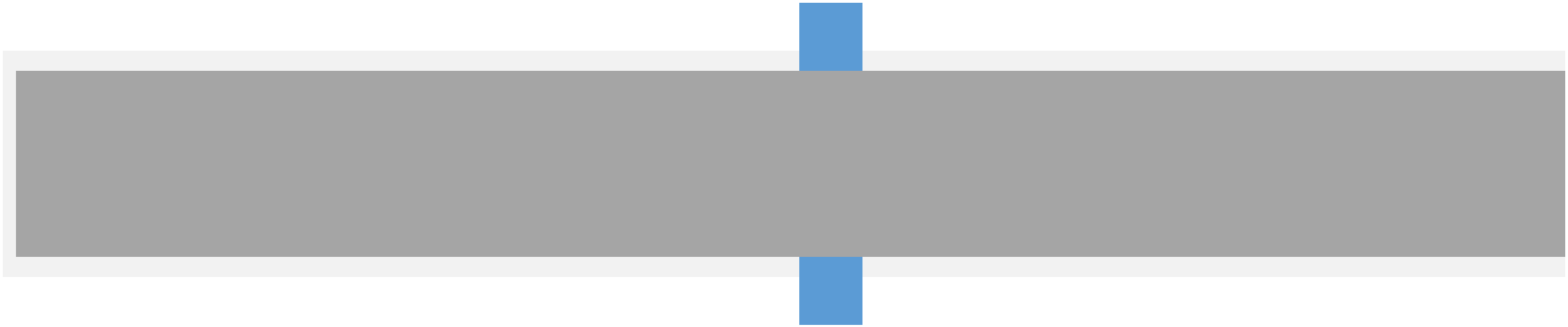}}\end{minipage}
                                    & {\color{gray} 1.89$\times$ }\\
MOON   								& 25 & \begin{minipage}[b]
                                    {0.14\columnwidth}\raisebox{-0.1\height}{
                                    \includegraphics[width=\linewidth]{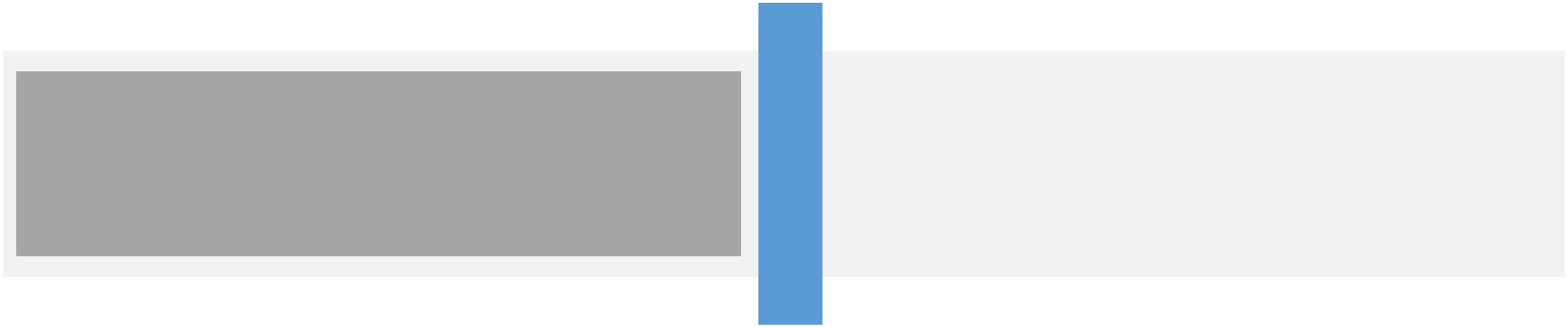}}\end{minipage}
                                    & {\color{gray} 0.89$\times$ }
                                    & 14 & \begin{minipage}[b]
                                    {0.14\columnwidth}\raisebox{-0.1\height}{
                                    \includegraphics[width=\linewidth]{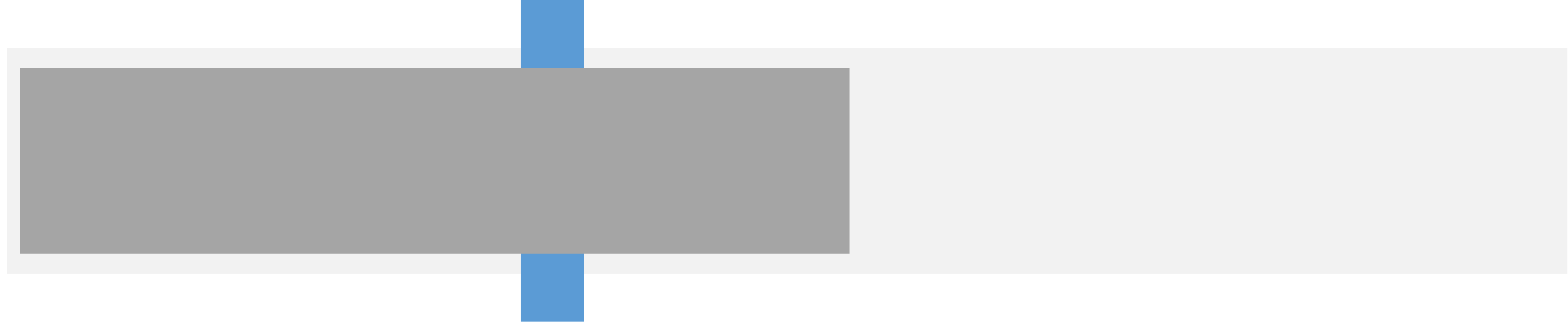}}\end{minipage}
                                     & {\color{gray} 1.56$\times$ }
                                    & 46 & \begin{minipage}[b]
                                    {0.14\columnwidth}\raisebox{-0.1\height}{
                                    \includegraphics[width=\linewidth]{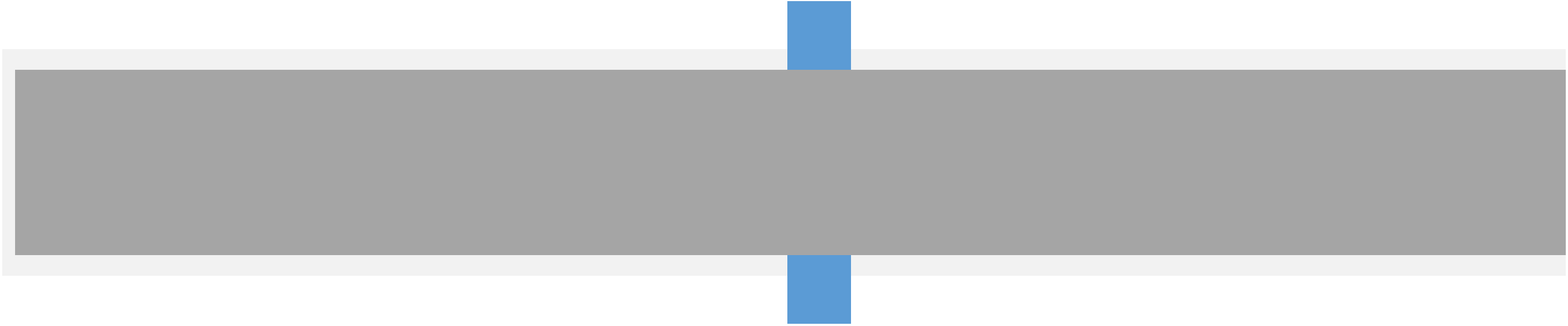}}\end{minipage}
                                    & {\color{gray} 1.92$\times$ }
                                    & 35 & \begin{minipage}[b]
                                    {0.14\columnwidth}\raisebox{-0.1\height}{
                                    \includegraphics[width=\linewidth]{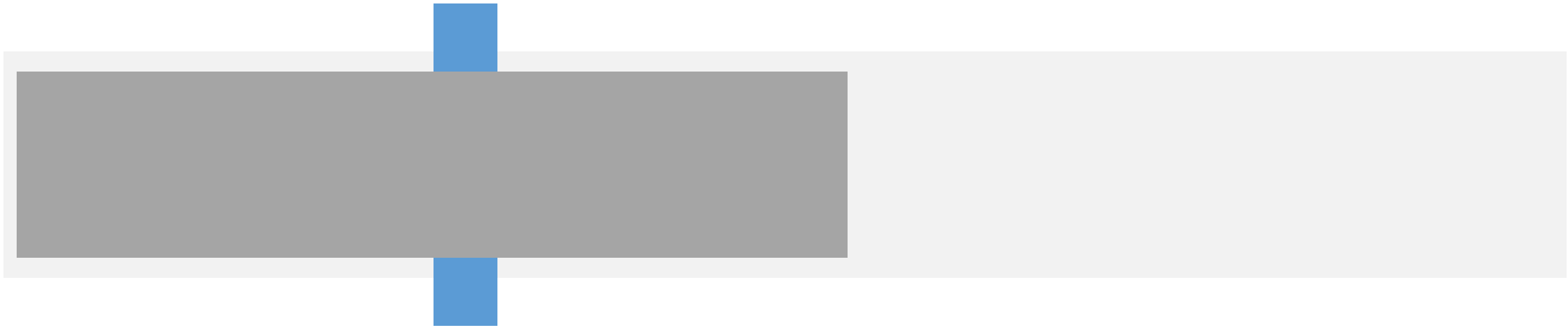}}\end{minipage}
                                    & {\color{gray} 1.84$\times$ }
                                    & 44 & \begin{minipage}[b]
                                    {0.14\columnwidth}\raisebox{-0.1\height}{
                                    \includegraphics[width=\linewidth]{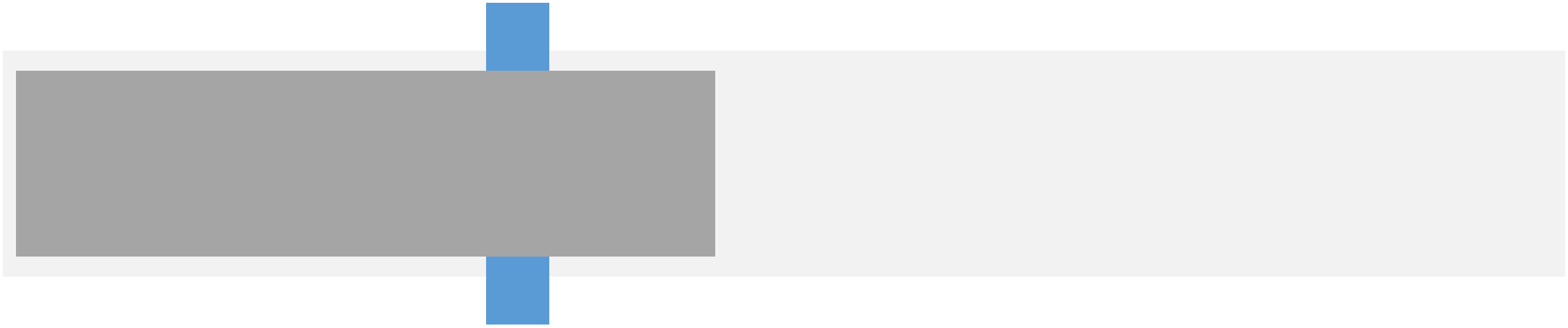}}\end{minipage}
                                    & {\color{gray} 1.38$\times$ }
                                    & 84 & \begin{minipage}[b]
                                    {0.14\columnwidth}\raisebox{-0.1\height}{
                                    \includegraphics[width=\linewidth]{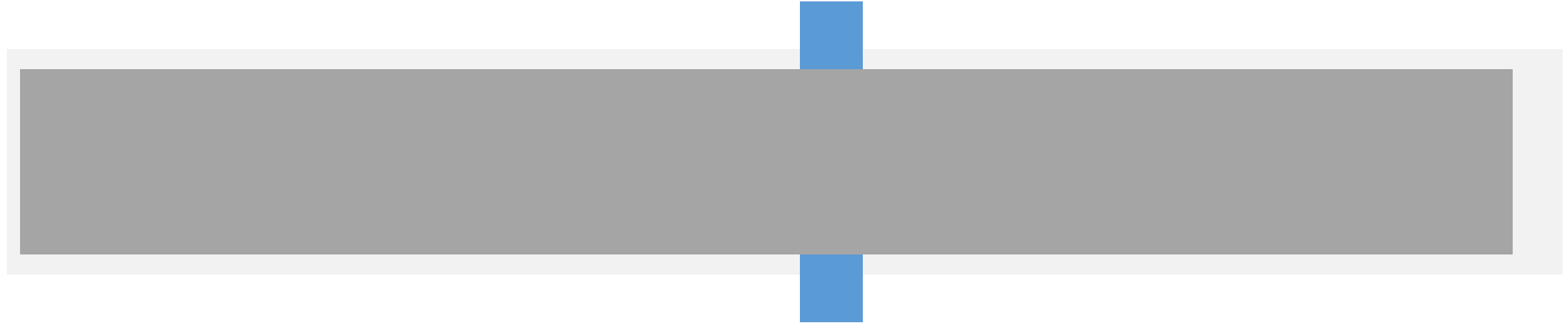}}\end{minipage}
                                    & {\color{gray} 1.75$\times$ }\\
FedDyn                              & 28 & \begin{minipage}[b]
                                    {0.14\columnwidth}\raisebox{-0.1\height}{
                                    \includegraphics[width=\linewidth]{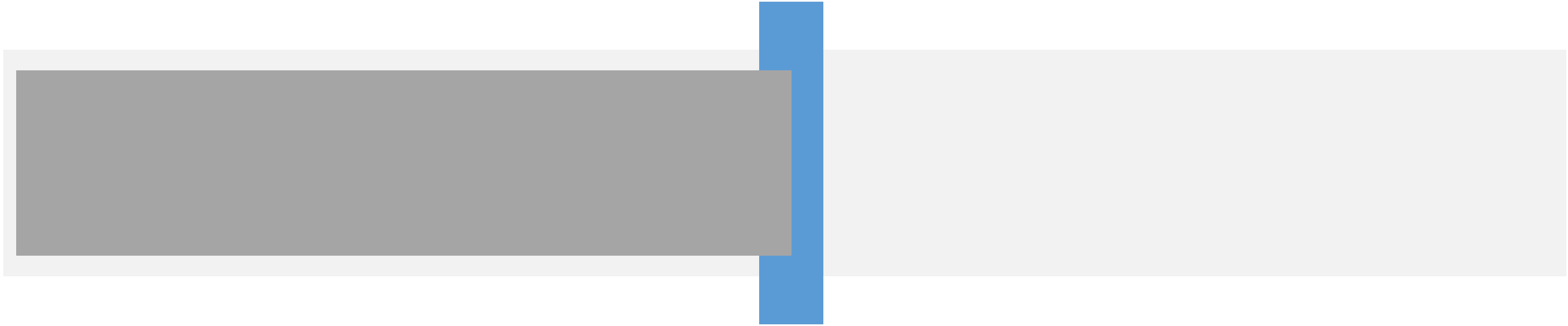}}\end{minipage}
                                    & {\color{gray} 1$\times$ }
                                    & 17 & \begin{minipage}[b]
                                    {0.14\columnwidth}\raisebox{-0.1\height}{
                                    \includegraphics[width=\linewidth]{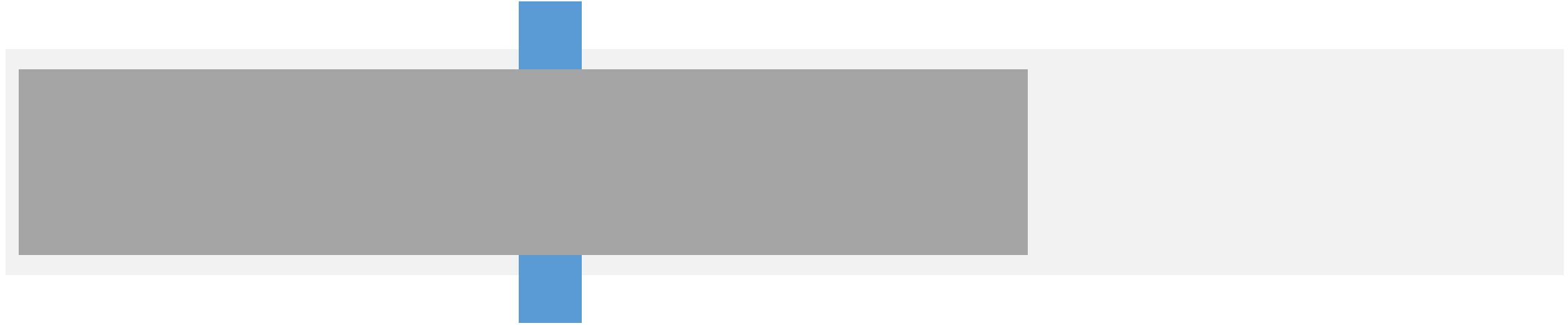}}\end{minipage}
                                    & {\color{gray} 1.89$\times$ }
                                    & 40 & \begin{minipage}[b]
                                    {0.14\columnwidth}\raisebox{-0.1\height}{
                                    \includegraphics[width=\linewidth]{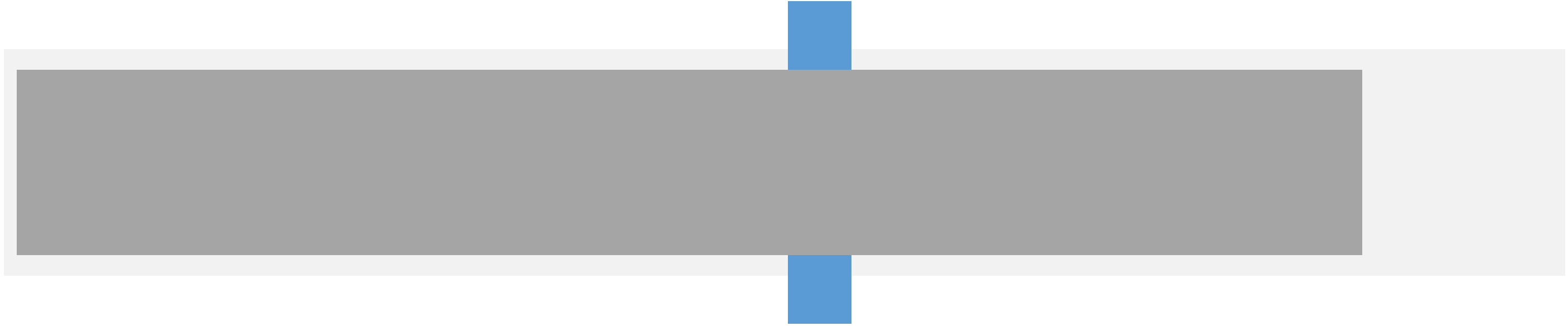}}\end{minipage}
                                    & {\color{gray} 2.08$\times$ }
                                    & 51 & \begin{minipage}[b]
                                    {0.14\columnwidth}\raisebox{-0.1\height}{
                                    \includegraphics[width=\linewidth]{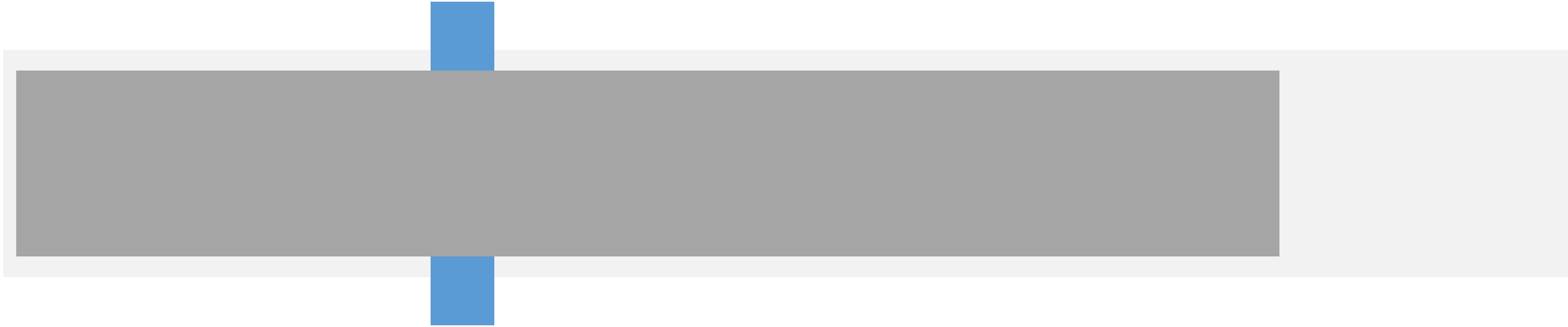}}\end{minipage}
                                    & {\color{gray} 2.68$\times$ }
                                    & 97 & \begin{minipage}[b]
                                    {0.14\columnwidth}\raisebox{-0.1\height}{
                                    \includegraphics[width=\linewidth]{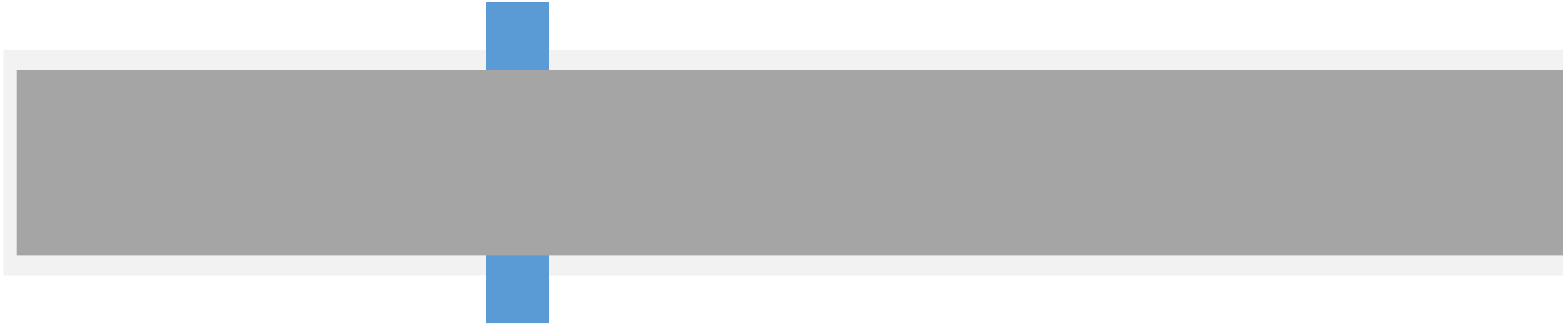}}\end{minipage}
                                    & {\color{gray} 3.03$\times$ }
                                    & 79 & \begin{minipage}[b]
                                    {0.14\columnwidth}\raisebox{-0.1\height}{
                                    \includegraphics[width=\linewidth]{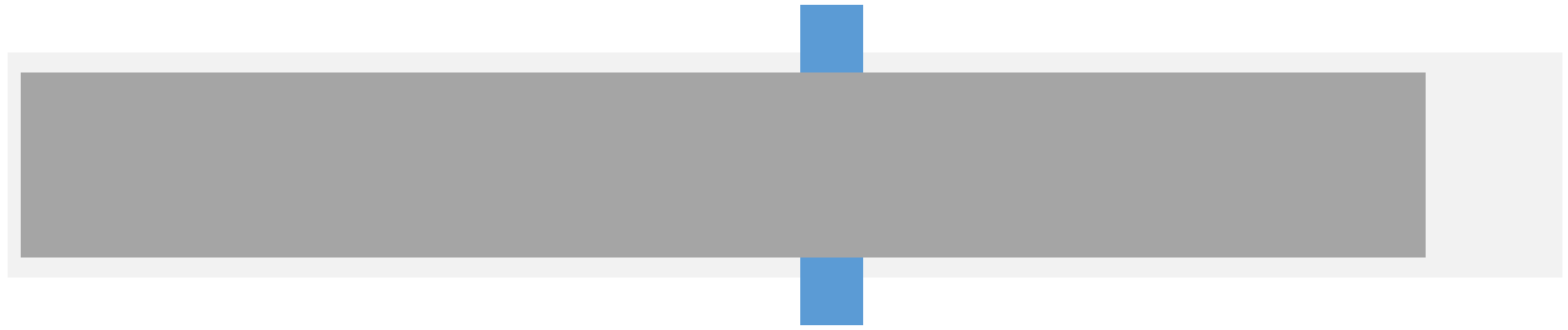}}\end{minipage}
                                     & {\color{gray} 1.72$\times$ }\\
\Xhline{2\arrayrulewidth}
\end{tabular}
% \vspace{-1.5em}
\end{table*}
Moreover, we simulate orthogonal data distribution, where clients are partitioned into multiple clusters. 
For each cluster, the data samples owned by inner clients have non-overlapped classes with those of other clusters, and the data samples of clients in each cluster are IID sampled. 
Concretely, we set two types of orthogonal data distribution by dividing clients into 5 and 10 clusters in our experiments, which are named $Orthogonal-\textit{5}$ and $Orthogonal-\textit{10}$ respectively.

Fig. \ref{fig_datapoints} shows the local data distributions of clients on MNIST dataset in 4 heterogeneity types.
The majority of clients contain mostly 3 or 4 classes of data samples under $Dir-\textit{0.5}$, and 1 or 2 classes of data samples under $Dir-\textit{0.5}$. 
Under $Orthogonal-\textit{5}$ and $Orthogonal-\textit{10}$, each client only has 2 and 1 classes of data samples. For example, Client 1 only have data samples with classes 0, 1 and class 0.\\
{\bf Baselines:} We compare the convergence performance of our proposed FedTrip with FedAvg \cite{mcmahan2017communication}, FedProx \cite{li2020federated}, SlowMo \cite{wang2019slowmo}, 
MOON \cite{li2021model} and FedDyn \cite{acar2021federated}. 
The default local optimizer is SGD with momentum (SGDm) \cite{sutskever2013importance}, a fixed learning rate of 0.01 and the momentum coefficient of 0.9.
Considering SGDm may results in performance degradation in some circustances, SlowMo and FedDyn choose SGD as the training optimizer. 
The hyperparameters of these methods are: $\mu=1.0$ for all MLP experiments and $\mu=0.4$ for others in FedTrip, $\mu=0.1$ in FedProx, $\alpha=1$ for the expertiments on MNIST dataset and $\alpha=0.1$ for other datasets in FedDyn, $\mu = 1, \tau=0.5$ in MOON. 
% Most of the hyperparameter values are consistent with the optimal value in their original paper.

The default number of communication rounds, batch size, and local epoch are set as 100, 50, and 1, respectively. 
Besides, the server randomly selects 4 devices from 10 devices.
We perform the 10-trial repeating experiment and report the average convergence performance.

\begin{figure*}[!thbp]
% \setlength{\abovecaptionskip}{-3pt}
% \subfigbottomskip=2pt
% \subfigcapskip=-4pt
  \begin{center}
    \subfigure[MNIST under $Dir-\textit{0.5}$]{\includegraphics[width=2.35in]{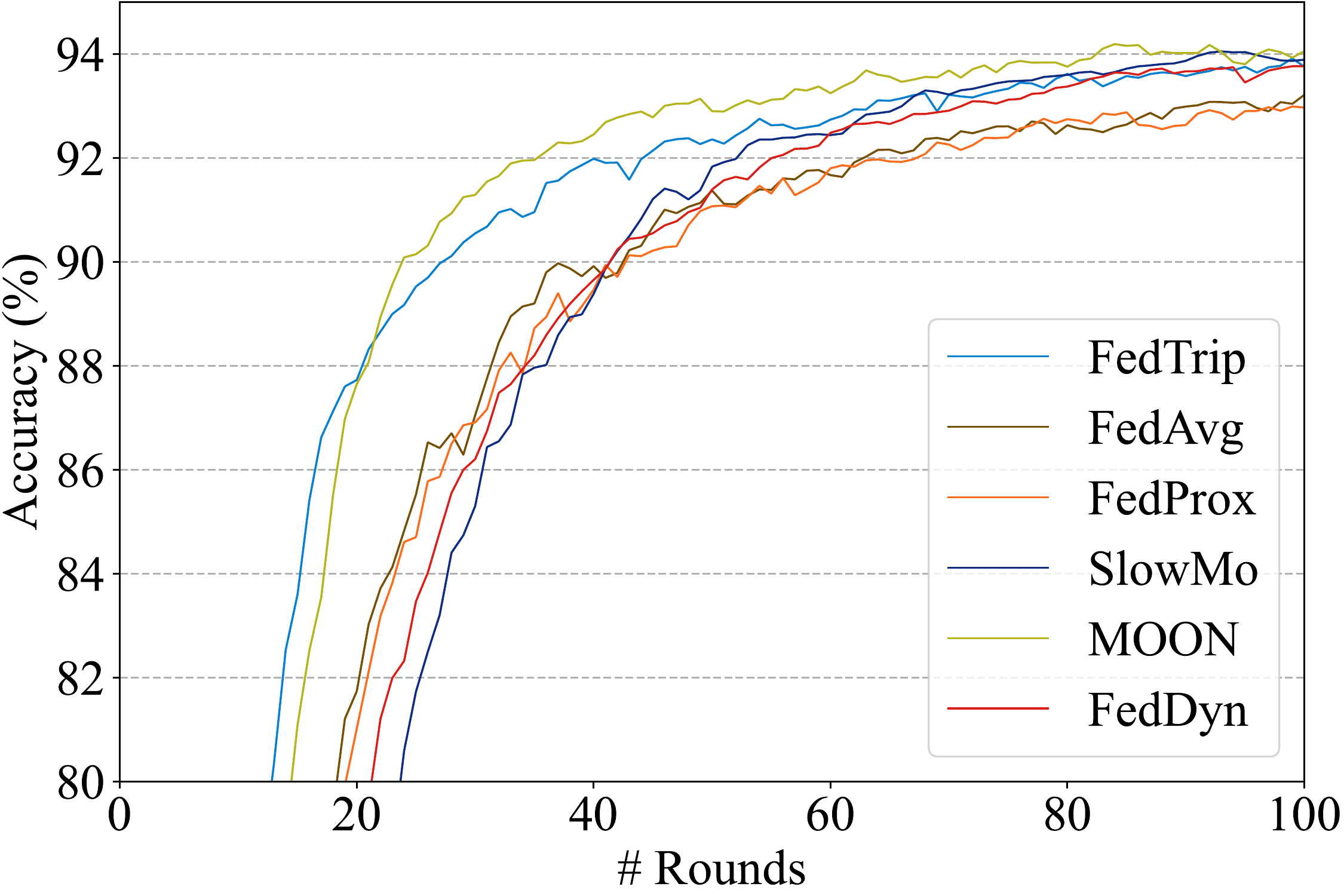}}
    \subfigure[FMNIST under $Dir-\textit{0.5}$]{\includegraphics[width=2.35in]{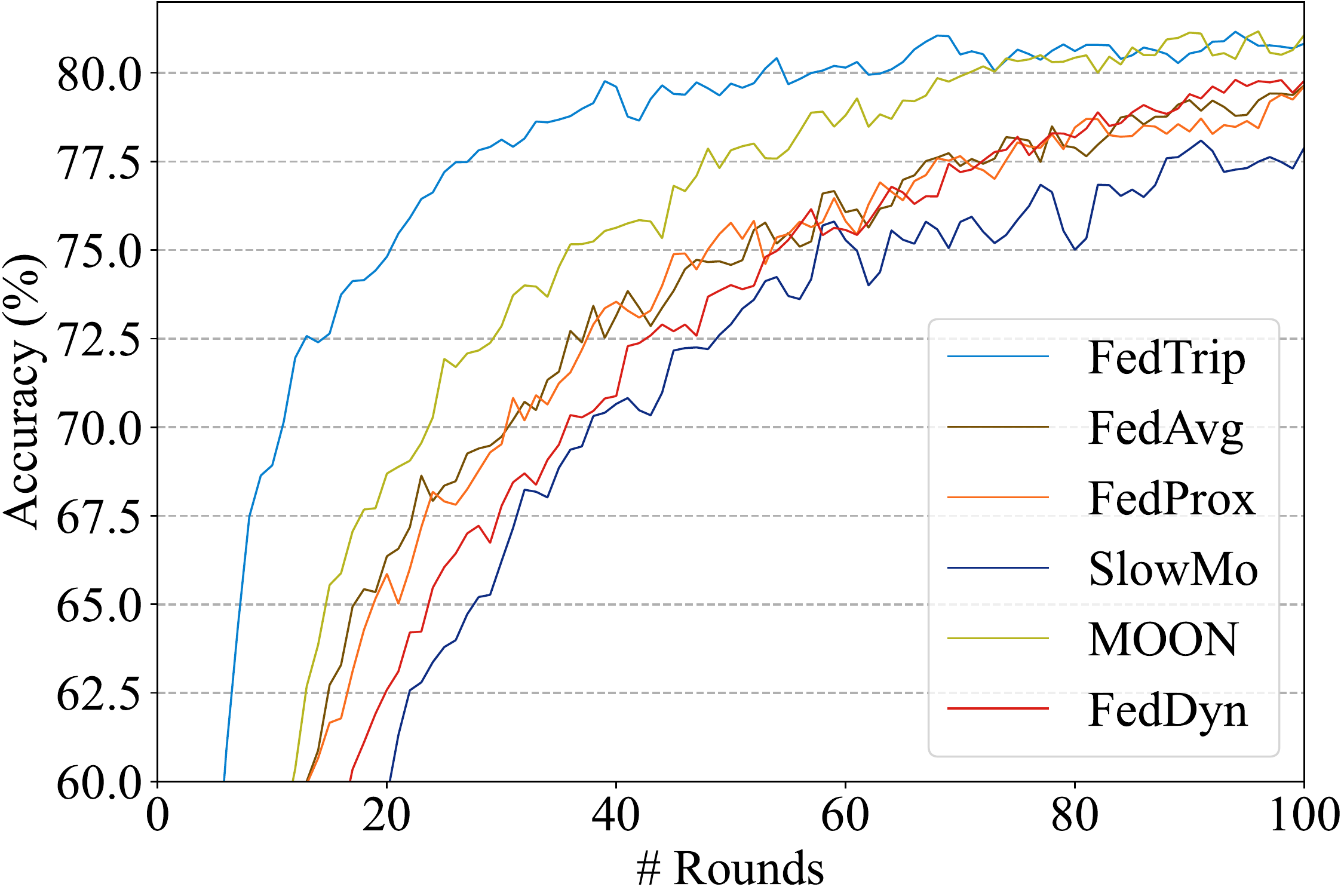}}
    \subfigure[EMNIST under $Dir-\textit{0.5}$]{\includegraphics[width=2.35in]{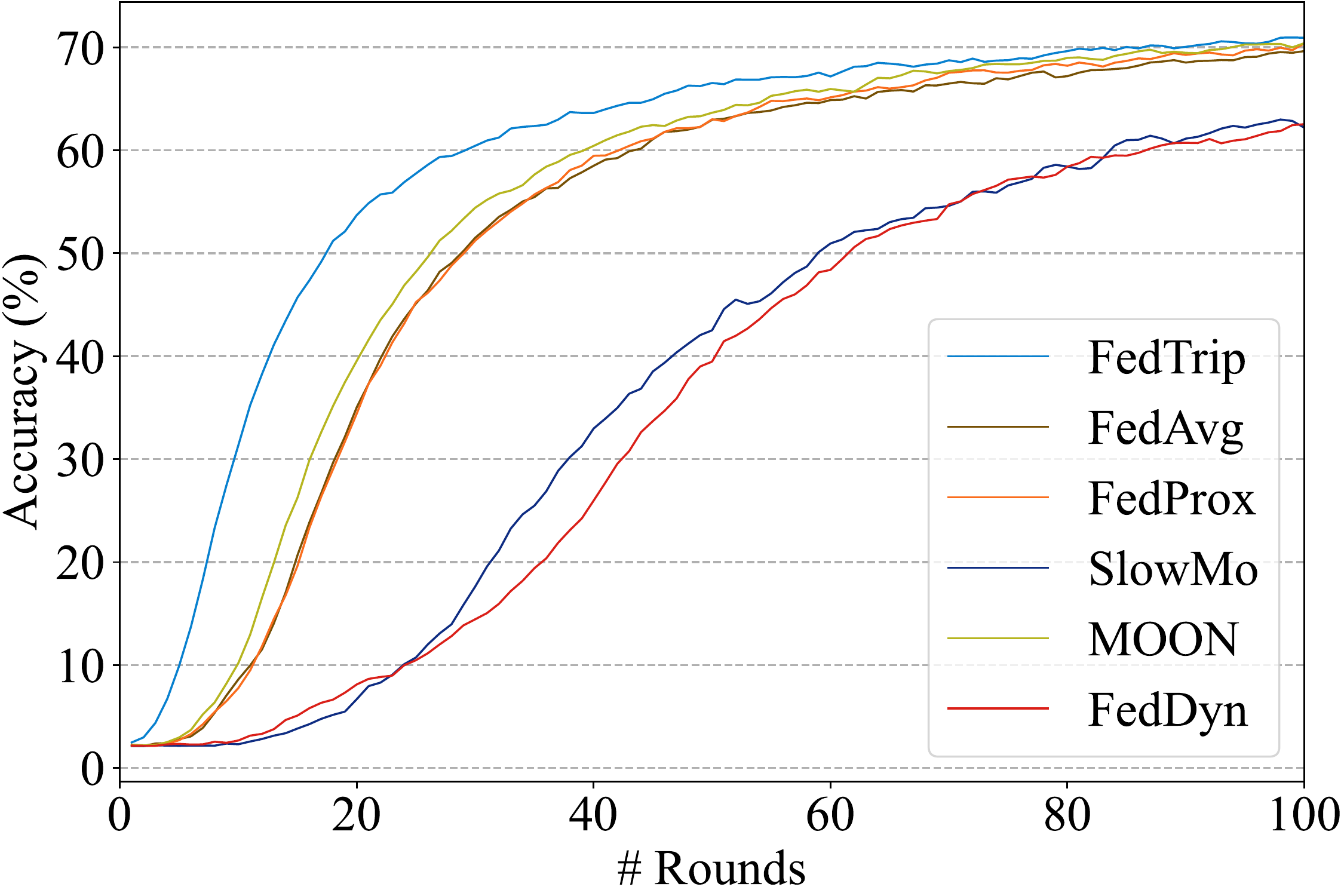}}
    \subfigure[MNIST under $Orthogonal-\textit{5}$]{\includegraphics[width=2.35in]{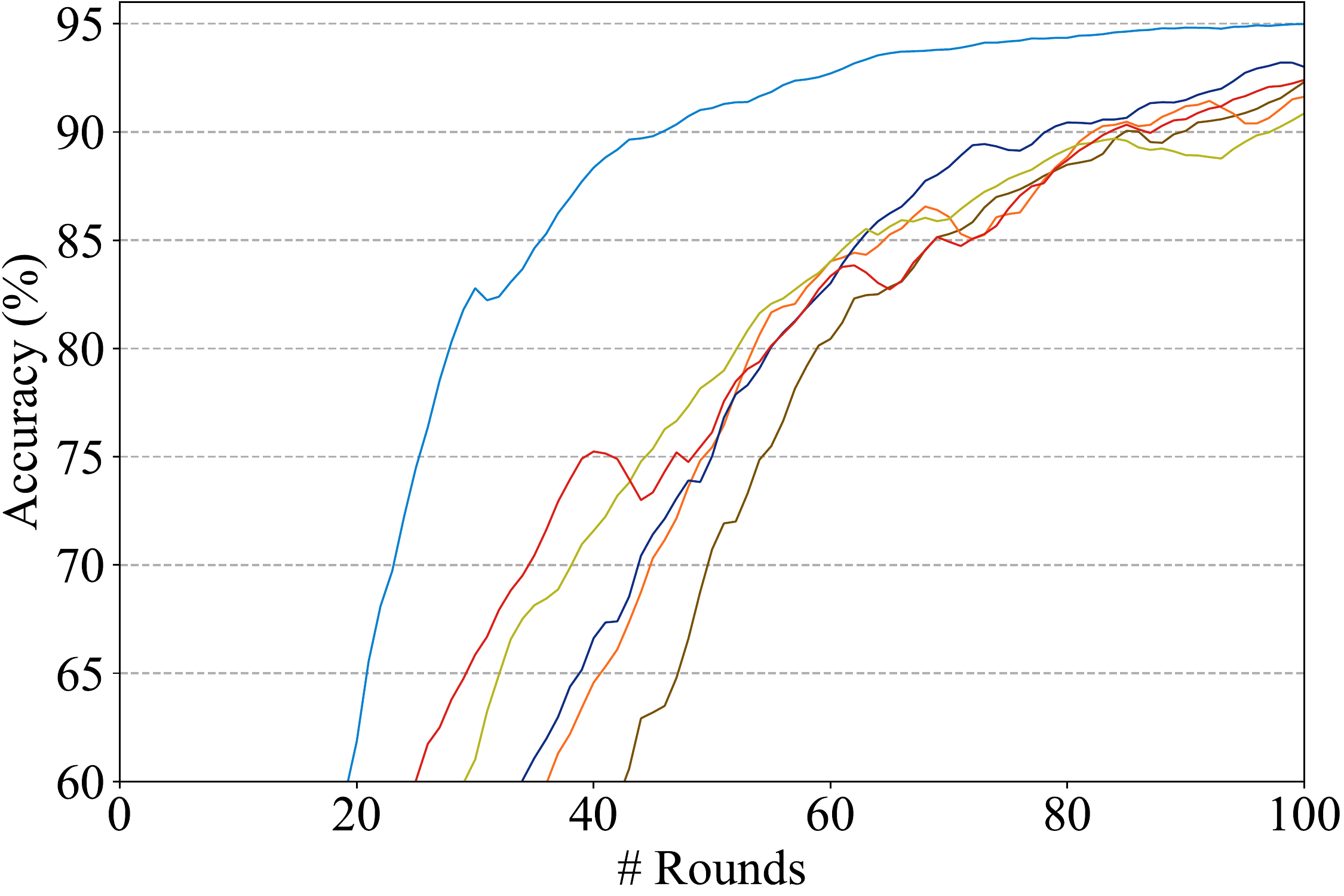}}
    \subfigure[FMNIST under $Orthogonal-\textit{5}$]{\includegraphics[width=2.35in]{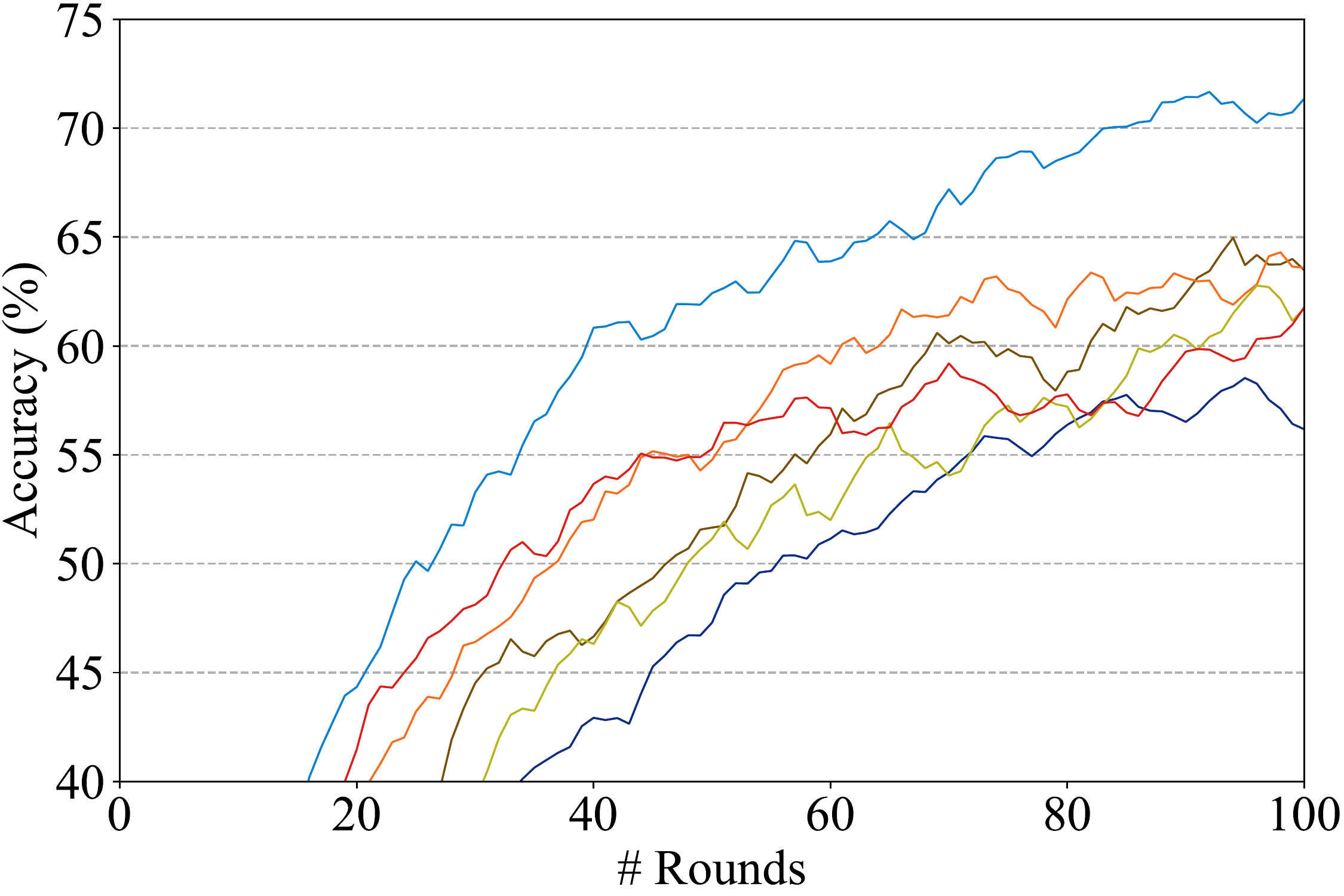}}
    \subfigure[EMNIST under $Orthogonal-\textit{5}$]{\includegraphics[width=2.35in]{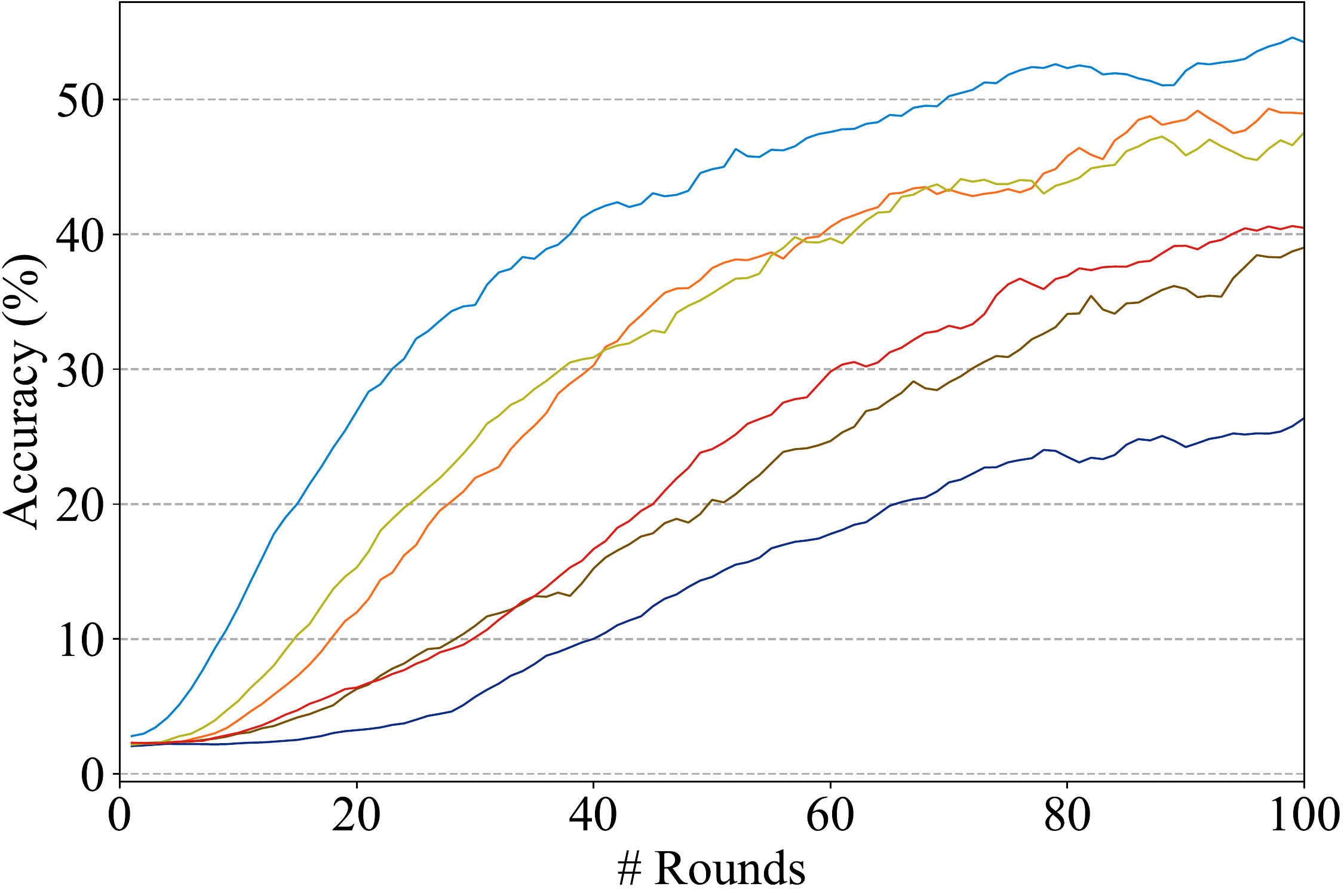}}
  \end{center}
  \caption{The convergence curves of CNN with 2 types of data heterogeneity on 3 datasets.}\label{fig_convergence}
  % \vspace{-1.5em}
\end{figure*}

\subsection{Resource Efficiency}
We verify the effectiveness of resource efficiency of FedTrip from the perspective of client-server communication and local computation. 
The results show that FedTrip is able to save resources significantly. \\
{\bf Communication Efficiency:}
As all aforementioned methods have exactly the same amount of communication volume per communication round, the total amount of communication bits is proportional to the number of communication rounds. 
We define the number of communication rounds at which the global model achieves the target accuracy as the evaluation metric. 
Table \ref{table_rate_10} shows the results of these methods on MLP, CNN, and AlexNet models under $Dir-\textit{0.5}$. 
The dark grey bars denote the number of communication rounds to achieve the target accuracy of the global model using different methods.
Thereinto, the longest dark grey bar indicates that the corresponding method has the maximum number of communication rounds.
Besides, the difference in the number of communication rounds between our proposed FedTrip and other methods can be shown by the blue lines. 

Among all the methods, FedTrip and MOON are the fastest. This demonstrates that absorbing information from both the global model and historical local models can effectively accelerate model convergence. Compared to MOON, FedTrip further reduces the communication rounds by 31.63\%, which shows that FedTrip absorbs model information more efficiently so as to further improve the convergence rate. 
Compared to the fundamental method FedAvg, FedTrip is 1.4-2.73$\times$ faster to achieve the target accuracy on training models, and the amount of communication rounds of FedTrip reduces by 44.02\% on average. 
We conclude that FedTrip shows the best performance on reducing communication overhead. \\
{\bf Local Computation Efficiency:}
Based on our theoretical analysis of computation cost over attaching operations in these methods (see Appendix \ref{FirstAppendix}), the computation cost of MOON is 50$\times$, 171.4$\times$ and 1,336$\times$ as much as that of FedTrip at each local iteration on training MLP, CNN and AlexNet, respectively.
We utilize the total GFLOPs of feedforward and attaching operations in these methods to measure computational efficiency, which are listed in Table \ref{FLOPS}.

\begin{table}[t]
\renewcommand{\arraystretch}{1.2}
    \setlength{\tabcolsep}{0.6mm}
    \caption{GFLOPs among methods during the training process}
    \centering
    \begin{tabular}{cccccccc}
    \Xhline{2\arrayrulewidth}
       {\bf Model} & {\bf Case} & {\bf FedTrip} & {\bf FedAvg} & {\bf FedProx} & {\bf SlowMo} & {\bf MOON} & {\bf FedDyn} \\
    \hline
          \multirow{2}{*}{MLP} 
           & MNIST & {\bf 1.441} & 2.334 & 2.626 & 2.191 & 3.573 & 1.441\\
           & FMNIST & {\bf 0.772} & 1.509 & 1.321 & 2.064 & 3.335 & 1.458\\
    \hline
          \multirow{3}{*}{CNN} 
           & MNIST & {\bf 6.161} & 9.897 & 10.465 & 10.151 & 35.02 & 10.269\\
           & FMNIST & {\bf 8.13} & 21.993 & 19.144 & 27.491 & 44.409 & 21,822\\
           & EMNIST & {\bf 41.077} & 57.097 & 57.431 & 116.733 & 167.486 & 124.513\\
    \hline
           AlexNet & CIFAR & {\bf 13,446} & 21,596 & 21,906 & 25,392 & 73,549 & 23,091\\
    \Xhline{2\arrayrulewidth}
    \end{tabular}
    \label{FLOPS}
% \vspace{-2em}
\end{table}

From Table \ref{FLOPS}, it can be seen that FedTrip reduces the computation cost by 39.58\% on average, compared to the baseline method with the least GFLOPs in each experiment case. 
The local computation overhead of MOON is 4.52$\times$ that of FedTrip, which demonstrates that FedTrip can obtain more convergence information with much less computation cost.
As MOON simultaneously obtains the information of the global and historical models via multiple feedforward operations, it is the most computation-inefficient method.
Compared to the fundamental method FedAvg, FedTrip reduces the local computation overhead by up to 42.27\%.

\begin{figure}[!t]
% \setlength{\abovecaptionskip}{-3pt}
% \subfigbottomskip=2pt
% \subfigcapskip=-4pt
  \begin{center}
    \subfigure[CNN]{\includegraphics[width=3.45in]{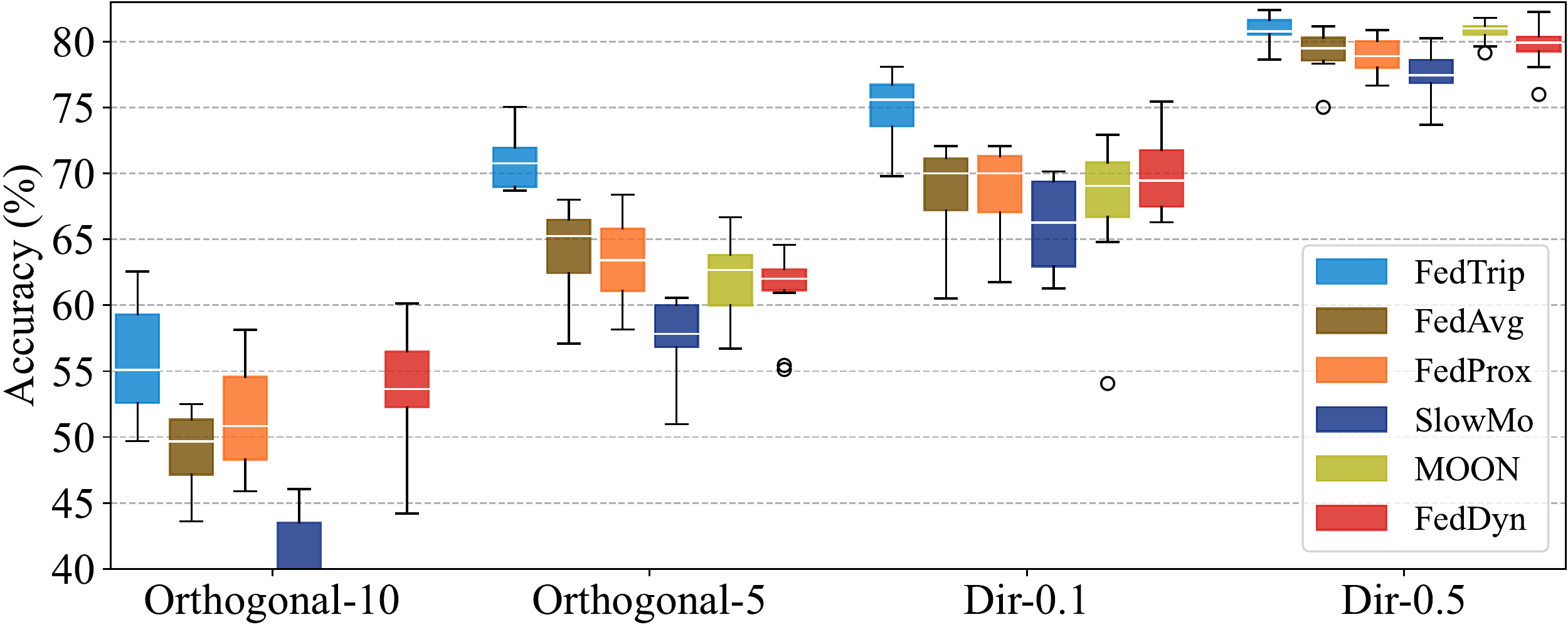}}
    \subfigure[MLP]{\includegraphics[width=3.45in]{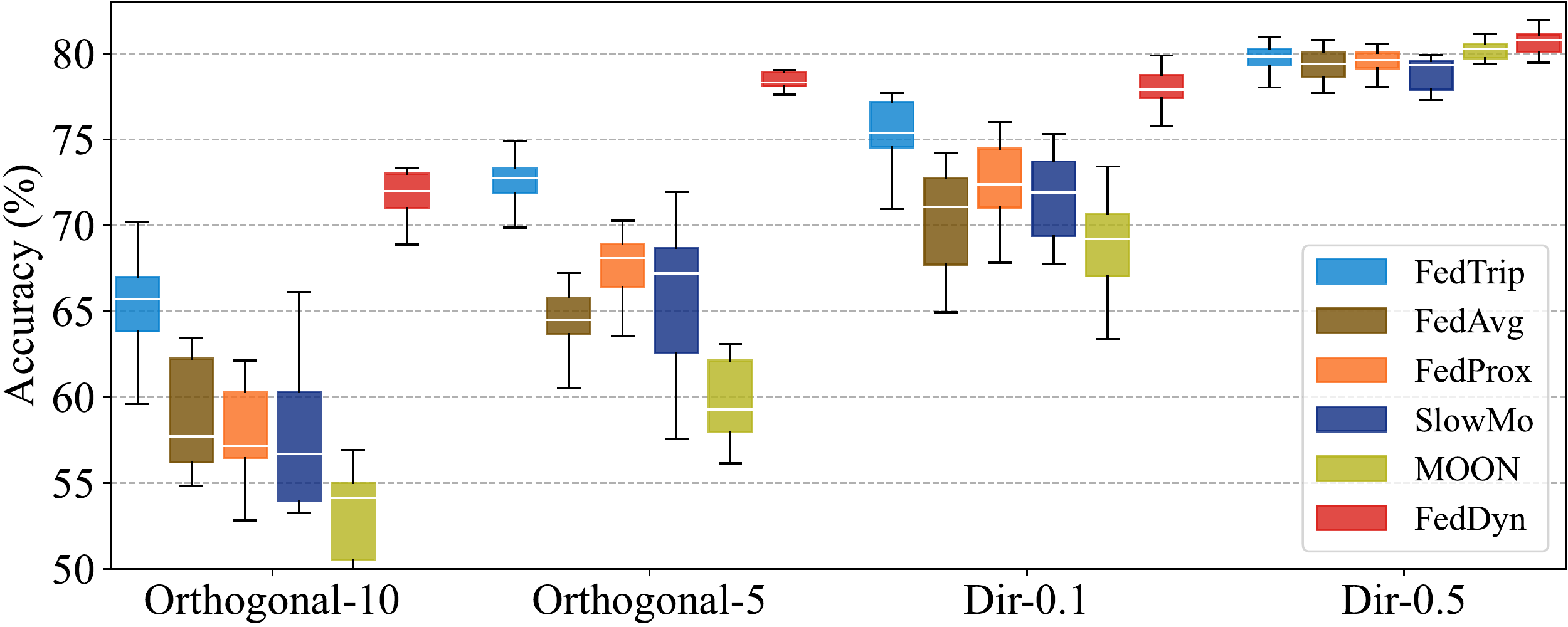}}
  \end{center}
  \caption{Boxplots of test accuracy of CNN and MLP with 4 types of data heterogeneity on FMNIST dataset. 
  The accuracy of MOON under $Orthogonal-\textit{10}$ is significantly lower than others, so it is invisible in (a).}\label{fig_heterogeneity}
  % \vspace{-2em}
\end{figure}

\subsection{Data Heterogeneity} 
Fig. \ref{fig_convergence} illustrates the test accuracy at each communication round for training CNN on different datasets among above methods. All the curves are smoothed by the exponential moving average.
As the value of $\mu$ is small in FedProx, the convergence performance of FedProx is generally close to that of FedAvg. 
As the regularization term in the local loss functions constrains the divergence of update directions among clients, FedProx becomes effective under orthogonal heterogeneity (see in Figs. \ref{fig_convergence}(e), \ref{fig_convergence}(f)), where the local updates diverge considerably. 
FedDyn and SlowMo underperform other methods on EMNIST dataset.
MOON \cite{li2021model} outperforms other baseline methods in $Dir-\textit{0.5}$, which indicates the advantages of absorbing the information of global and historical models. 
FedTrip is competitive to data heterogeneity as it outperforms other baselines in most of experiment cases. 

Fig. \ref{fig_heterogeneity} illustrates the final accuracy, the average accuracy of the global model over the last 10 communication rounds of CNN and MLP on FMNIST dataset among all methods. FedTrip performs the highest final accuracy in most of the experiments under various heterogeneity types. 
FedDyn has the highest accuracy on MLP under the orthogonal heterogeneity types, which shows its superiority of the related regularization term for aligning the local model and local gradients. 
Compared to the experimental results under the Dirichlet distribution, the convergence improvement of FedTrip is more remarkable than the improvement in the experiments under the orthogonal distribution.
Although MOON can simultaneously obtain the global model and historical model information, it conducts the worst performance in $Orthogonal-\textit{10}$. 
This reveals that this model representation method is not suitable for all data distributions, especially highly-skewed data distributions.
Overall, FedTrip achieves 2.53$\times$ and 1.38$\times$ convergence acceleration compared to the state-of-art method MOON in $Dir-\textit{0.1}$ and $Orthogonal-\textit{10}$ respectively. 
We attribute this to that FedTrip ables to mitigate the server fluctuation in the convergence trajectory in heavily-skewed data distributions.

\subsection{Scalability} 
We discuss the scalability of FedTrip based on the client participation type that the server randomly selects 4 devices from 50 devices.
The convergence performances across different models and data heterogeneity types are listed in Table \ref{table_rate_CNN_50}. Symbol $>$ in Table \ref{table_rate_CNN_50} indicates that the global model of the specific method does not achieve the target accuracy at the last communication round. 

The communication rounds of these methods to achieve the target accuracy in $\textit{4-50}$ are less than that in $\textit{4-10}$ with the same hyperparameters, which is benefit from the larger number of total data samples.
With the details of FedTrip in Section \Rmnum{4}, 
to scale the influence of the historical local model,
$\xi$ in FedTrip is scaled by the gap between the current round and the last participated round. The expectation value of $\xi$ decreases to $\frac{1}{5}$ of $\xi$ in $\textit{4-10}$. 
Among experiments, FedTrip performs the fastest convergence in $\textit{4-50}$. Compared to FedAvg and MOON, FedTrip reduces communication rounds by up to 56.1\% and 54.82\% respectively. 
The performance of MOON degrades in this setting, which shows the limitation of MOON in low client participation environments.
In summary, FedTrip consistently yields substantial resource savings compared to baselines across various client participation settings.

\begin{table}[t]
% \setlength{\abovecaptionskip}{0.cm}
\renewcommand{\arraystretch}{1.2}
\setlength{\tabcolsep}{1mm}
\caption{The Number of communication rounds of CNN to achieve the target accuracy in $4-50$.}
\label{table_rate_CNN_50}
\centering
\begin{tabular}{ccccccc}   
\Xhline{2\arrayrulewidth}
 \multirow{3}{*}{{\bf Method}} 
& \multicolumn{3}{c}{{\bf MNIST}} & \multicolumn{3}{c}{{\bf FMNIST}} \\
\cline{2-7}
& Dir-0.1 & Dir-0.5 & Orthogonal-5 & Dir-0.1 & Dir-0.5 & Orthogonal-5 \\
& 87\% & 90\% & 85\% & 65\% & 75\% & 60\% \\
\hline
FedTrip 							& {\bf 30} & {\bf 19}  & {\bf 43}
                                    & {\bf 19} & {\bf 15}  & {\bf 35}\\
FedAvg     					        & 1.6$\times$ & 1.74$\times$ & 2.14$\times$
                                    & 2.74$\times$ & 3$\times$ & 2.51$\times$\\
FedProx 						    & 1.8$\times$ & 1.71$\times$ & 1.7$\times$
                                    & 2.68$\times$ & 2.87$\times$ & 2.14$\times$\\
SlowMo  						    & 1.87$\times$ & 1.71$\times$ & 1.7$\times$
                                    & 4.21$\times$ & 4.67$\times$ & $>$2.86$\times$\\
MOON                                & 2.33$\times$ & 1.32$\times$ & 2.28$\times$ 
                                    & 4$\times$ & 2.67$\times$ & $>$2.86$\times$ \\
FedDyn                              & 2.17$\times$ & 3$\times$ & 2.28$\times$
                                    & 4.16$\times$ & 5.07$\times$ & $>$2.86$\times$\\
\Xhline{2\arrayrulewidth}
\end{tabular}
% \vspace{-1em}
\end{table}
\par

\subsection{Influence of Aggregation Intervals}
In this part, we enlarge the number of local training epochs to 5 and 10 at each communication round. 
The experiments run under settings with $Dir-\textit{0.5}$ and $\textit{4-10}$. We set $\mu=0.4$ in FedTrip. 
We list the test accuracy of each method at the 10-th and 20-th communication round in Table \ref{tab:local epochs}.
FedTrip consistently achieves the highest accuracy among different aggregation intervals. 
With the increase of local training iterations per communication round, the average accuracy of all methods at each round is improved. 
Although a large aggregation interval exacerbates the staleness of historical local models, our method can still obtain the effective information from the historical models to accelerate convergence. 
SlowMo and FedDyn have unsatisfactory performance owing to the frequency reduction of the additional operations at the server, resulting in incorrect updates.

\begin{table}[t]
\renewcommand{\arraystretch}{1.2}
    \setlength{\tabcolsep}{0.7mm}
    \caption{The accuracy among methods with 5 and 10 of local epochs.}
    \centering
    \begin{tabular}{cccccccc}
    \Xhline{2\arrayrulewidth}
       \# {\bf Local} & \multirow{2}{*}{\#{\bf Rounds}} & \multirow{2}{*}{{\bf FedTrip}} & \multirow{2}{*}{{\bf FedAvg}} & \multirow{2}{*}{{\bf FedProx}} & \multirow{2}{*}{{\bf SlowMo}} & \multirow{2}{*}{{\bf MOON}} & \multirow{2}{*}{{\bf FedDyn}} \\
       {\bf Epochs} &  &  &  &  &  &  \\
    \hline
        \multirow{2}{*}{5}  & 10 & {\bf 96.36} & 95.49 & 93.08 & 84.55 & 95.26 & 87.93\\
                            & 20 & {\bf 97.18} & 96.71 & 95.95 & 92.88 & 96.88 & 93.49\\
        \multirow{2}{*}{10} & 10 & {\bf 97.49} & 97.38 & 95.84 & 87.79 & 96.99 & 93.11\\
                            & 20 & {\bf 97.95} & 97.84 & 97.25 & 95.15 & 97.84 & 95.93\\
    \Xhline{2\arrayrulewidth}
    \end{tabular}
    \label{tab:local epochs}
    % \vspace{-1.5em}
\end{table}

\subsection{Sensitivity Analysis of $\mu$ in FedTrip}
\begin{figure}[!t]
% \setlength{\abovecaptionskip}{-3pt}
% \subfigbottomskip=2pt
% \subfigcapskip=-4pt
    \subfigure[CNN on MNIST under $Dir-0.1$]{{\includegraphics[width=\linewidth]{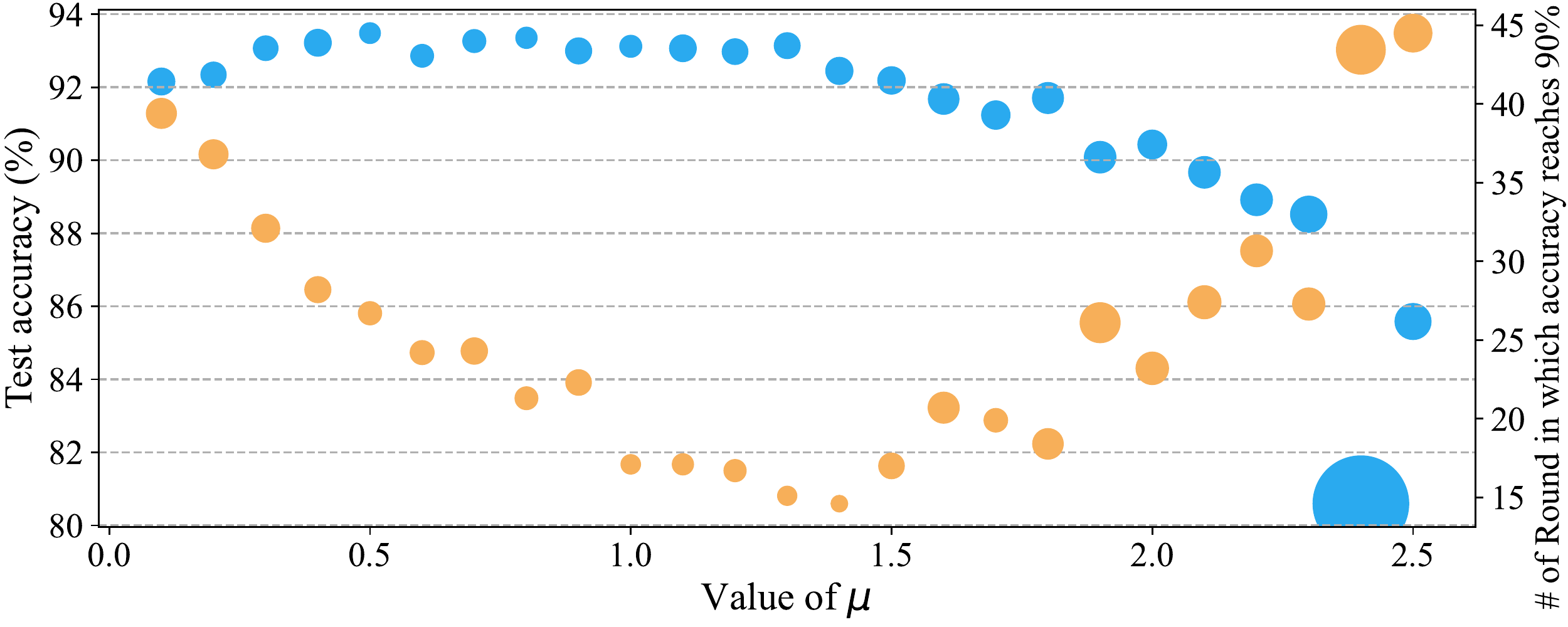}}}
    \subfigure[CNN on MNIST under $Dir-0.5$]{{\includegraphics[width=\linewidth]{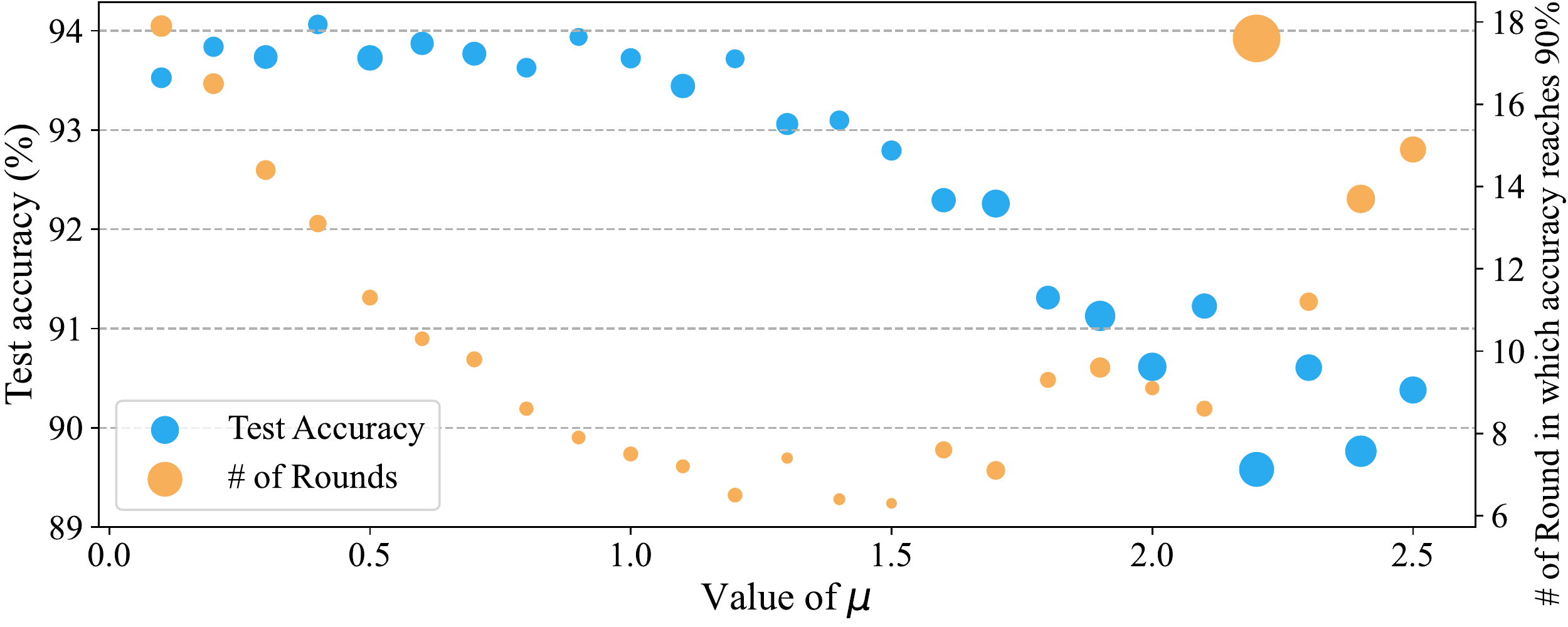}}}
    \subfigure[CNN on MNIST under $Orthogonal$]{{\includegraphics[width=\linewidth]{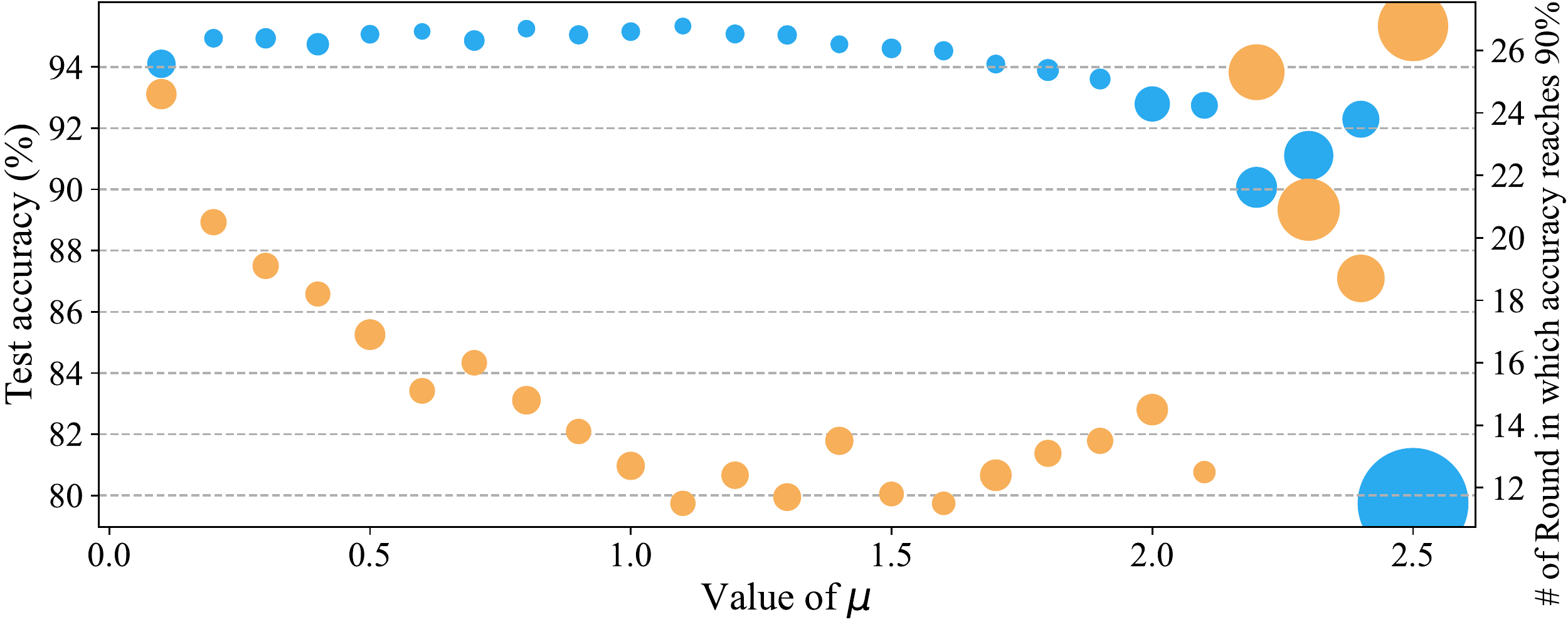}}}
    \subfigure[MLP on FMNIST under $Dir-0.5$]{{\includegraphics[width=\linewidth]{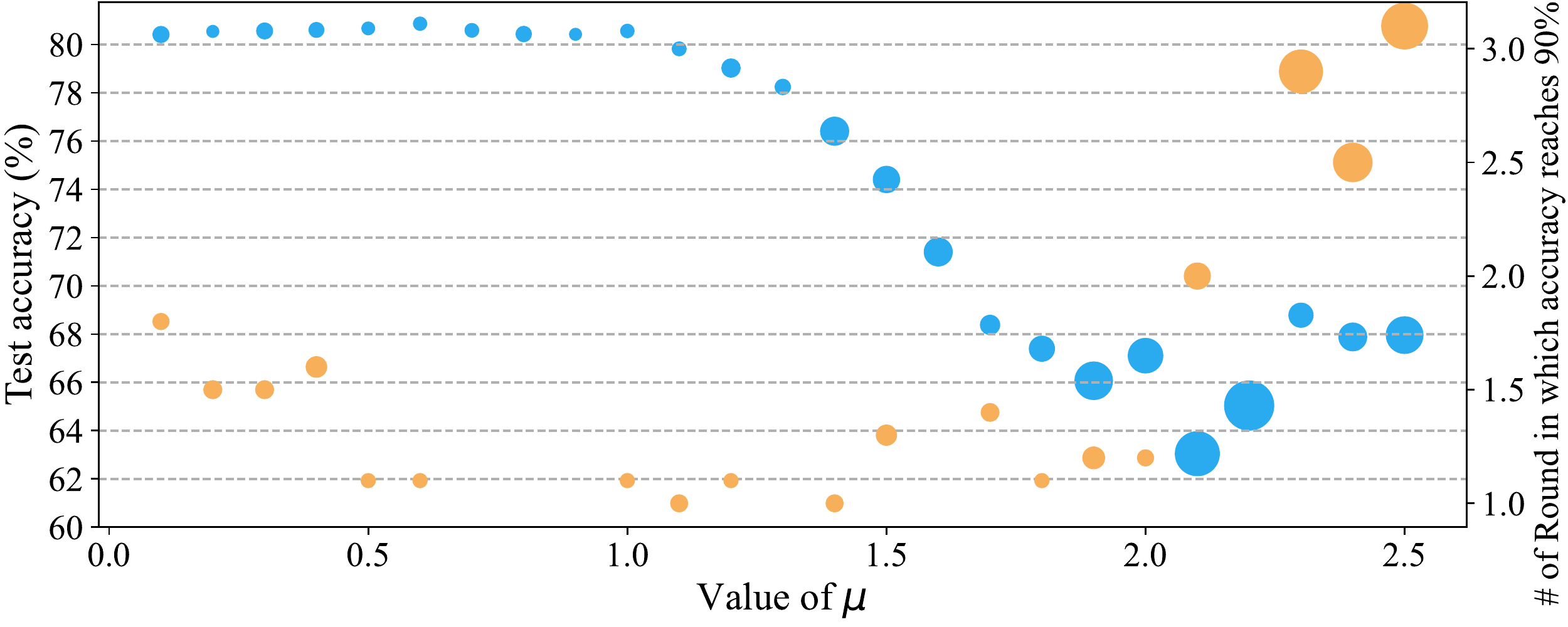}}}
  \caption{The performance influence of $\mu$ on CNN model and MNIST and MLP model on FMNIST.}\label{fig_mu}
  % \vspace{-1.5em}
\end{figure}

To explore the influence of $\mu$ on the convergence, we compare the model accuracy and convergence rate of FedTrip by varying $\mu$ from 0.1 to 2.5. 
Note that the final accuracy is defined as the highest test accuracy in the training process, which indicates the best performance across different values of $\mu$.  
The model, dataset, and participation type are CNN, MNIST, and $\textit{4-10}$ respectively. The results are shown in Fig. \ref{fig_mu}.
The blue and the orange circles represent the final accuracy and the number of communication rounds required to achieve the 90\% test accuracy of the global model. 
Note that, the radii of circles represent the variance of corresponding metrics.

Under all settings, FedTrip eventually converges successfully. 
As shown in Fig. \ref{fig_mu}, FedTrip suffers a lower convergence rate when the value of $\mu$ is small. 
It accelerates convergence and improves test accuracy to 93.48\% and 94.06\% under $Dir-0.1$ and $Dir-0.5$ when $\mu=0.4$. 
Afterward, the convergence is still accelerated at the sacrifice of accuracy degradation when the value of $\mu$ increases to approximately 1.5. 
With further increasing $\mu$, the number of communication rounds increases, and the final accuracy decreases. 
Under $Dir-0.1$, the test accuracy of FedTrip fluctuates more considerably and degrades faster with increasing $\mu$ than that under $Dir-0.5$. Under $Orthogonal$ setting, our method has the more stable performance than that under Dirichlet data heterogeneity with the change of $\mu$. The performance fluctuates dramatically When $\mu > 2$, but the test accuracy drops to 80\% only when $\mu=2.5$. As shown in Figure \ref{fig_mu} (d), the performance is sensitive to $\mu$. With the value of $\mu$ increases, the test performance considerably degrades.

Consequently, for devices with limited resource budgets or with less-strict performance requirements, a large $\mu$ is a better option. 
Conversely, we need to set a small $\mu$ for the cases with high performance requirements.

\section{Conclusion}
In this paper, we propose a resource-efficient FL method named FedTrip based on the insight of constraining local model close to the global model while keeping away from the historical model, aiming to mitigate the impact of inconsistency model update derived from data heterogeneity and effectively obtain the information that helps fast convergence.
Specifically, we add a triplet regularization term to the client-side loss function to absorb the useful information from historical local model with negligible computation cost. 
Our experiments show that FedTrip outperforms the state-of-the-art methods in terms of reducing computation and communication overhead under the circumstances of varying data heterogeneity, client participation and hyperparameter settings. 
In future, we will further discuss the influence of $\xi$ and analyze the convergence rate of FedTrip in general convex and non-convex cases.

\appendices
\section{Discussion on resource consumption of attaching operations in methods}
\label{FirstAppendix}

\begin{table}[t]
\renewcommand{\arraystretch}{1.2}
\setlength{\tabcolsep}{1.2mm}
\caption{Comparison of FedTrip with related state-of-the-arts on the computation and communication overhead.}
\label{table_comparision}
\centering
\begin{tabular}{ccc}
\Xhline{2\arrayrulewidth}
\multirow{2}{*}{\textbf{Method}} & \textbf{Computation} & \textbf{Communication}\\
 & \textbf{Overhead (FLOPs)} & \textbf{Overhead}\\
\hline
 SCAFFOLD \cite{karimireddy2020scaffold} & $2(K+1)\lvert w\rvert+ n(FP + BP)$ & $2\lvert w \rvert$ \\
 MimeLite \cite{karimireddy2020mime} & $n(FP + BP)$ & $2\lvert w \rvert$  \\
 MOON \cite{li2021model }& $K\left(M(1+p)FP\right)$ & $0$  \\
 FedProx \cite{li2020federated} & $2K\lvert w\rvert$ & $0$\\
 FedDyn \cite{acar2021federated} & $4K\lvert w\rvert$ & $0$\\
 FedTrip & $4K\lvert w\rvert$ & $0$  \\
\Xhline{2\arrayrulewidth}
\end{tabular}
\label{tab:computation}
% \vspace{-1.5em}
\end{table}

This part illustrates the computation and communication consumption of FedTrip compared with related state-of-the-art methods. 
FedAvg is the baseline, and all methods use the SGDm optimizer. 
Table \ref{table_comparision} displays the consumption at each communication round of attaching operations in these methods.
We define $K, M, n, \lvert w\rvert$ as the number of local iterations, batch size, local data samples, and the size of model parameters.
In addition, $\mathcal{FP}$ and $\mathcal{BP}$ represent the computation overhead of feedforward and backpropagation for a single data sample, respectively. 

SCAFFOLD and MimeLite require extra computation of full-batch gradients to estimate the true gradients, which requires $n(\mathcal{FP}+BP)$ computation overhead. 
Note that the computation overhead of $\mathcal{FP}$ and $\mathcal{BP}$ is much larger than $\lvert w\rvert$. For example, the computation overhead of $\mathcal{FP}$ is 342 $\times$ as much as $\lvert w\rvert$ on CNN model.
In addition, they require extra transmission in both downstream and upstream communication. The size of transmission is $2\lvert w\rvert$. 
MOON requires extra $(1+p)\mathcal{FP}$ for each local iteration, where $p$ denotes the number of history models used in the local iterations. 
FedTrip only requires $4\lvert w\rvert$ computation overhead at each local iteration, much smaller than that of MOON. 

In general, although existing methods can alleviate data heterogeneity and improve convergence performance, they require a large amount of local computation or client-server interactions because of their attaching operations. 
The attaching operations of FedTrip not only doesn't increase additional communication overhead, but its computation overhead is much smaller than that of other methods, almost negligible. 
% FedTrip is the best method, especially in resource-constrained environments.

\section{Sketch of convergence proof}
\label{SecondAppendix}
We prove the convergence analysis of FedTrip by referring to the proof of FedProx \cite{li2020federated} and FedDANE \cite{li2019feddane}. Besides, one lemma is adapted from Scaffold \cite{karimireddy2020scaffold}, which we will apply in $\langle \nabla f(w^t), w^{t+1}-w^t\rangle$:
\begin{align}
    \langle \nabla f(x), y-z) \leq f(y) - f(z) - \frac{\mu}{4}\lVert y - z \rVert^2 + L \lVert z - x \rVert^2.    
\end{align}
First, we define $e_k^t$ such that:
\begin{align}
	\notag \nabla F_k (w_k^{t+1}) + & \mu(w_k^{t+1} - w^t) + \mu\xi_k^t(\tilde{w}_k^t - w^{t+1}_k) = e_k^t\\
    \notag &\lVert e_k^t\rVert\leq \gamma\lVert \nabla F_k(w^t)\rVert,
\end{align}
and we define $\mathbb{E}_k[w_k^{t+1}]=\bar{w}^{t+1}$.
We have
\begin{align}
    \notag (\mathbb{E}_k[w_k^{t+1}] - w^t) + & \mathbb{E}_k[\xi_k^t(\tilde{w}_k^t - w_k^{t+1})] \\
    &= -\frac{1}{\mu}\mathbb{E}_k[\nabla F_k (w_k^{t+1})]+\frac{1}{\mu} \mathbb{E}_k[e_k^t].
\end{align}
Then we set $\hat{w}^{t+1}_k=arg\min_w h_k(w;w^t)$. It's obviously that$\nabla h_k(\hat{w}^{t+1}_k,w^t) = 0$. Due to the $\mu$-strongly convex of $h_k$, we can get
\begin{align}
    \notag \lVert \hat{w}^{t+1}_k - w^{t+1}_k \rVert 
    & = \frac{1}{\mu}\lVert\nabla h(\hat{w}^{t+1}_k,w^t) - \nabla h( w^{t+1}_k,w^t)\rVert\\
  \notag   & = \frac{1}{\mu}\lVert e_k^t \rVert
    \ \leq\ \frac{\gamma}{\mu}\lVert\nabla F_k(w_k^t) \rVert,\\
    \notag \lVert \hat{w}^{t+1}_k - w^{t} \rVert
     & = \frac{1}{\mu}\lVert\nabla h(\hat{w}^{t+1}_k,w^t) - \nabla h( w^t,w^t)\rVert\\
  \notag  & = \frac{1}{\mu}\lVert - \nabla F_k(w^t) - \mu\xi_k^t( \tilde{w}_k^t - w^t)\rVert,\\
    \lVert  w^{t+1}_k - w^{t} \rVert  
    & \leq \frac{1+\gamma}{\mu}\lVert \nabla F_k(w^t) \rVert + \xi_k^t\lVert\tilde{w}_k^t - w^t\rVert.
\end{align}
As $\mathbb{E}_k [\lVert  w^{t+1}_k - w^{t} \rVert ]  = \lVert \bar{w}^{t+1} - w^t\rVert$, we use Assumption 1 to get
\begin{align}
	\notag\lVert &\bar{w}^{t+1} - w^t\rVert \leq \frac{B(1+\gamma)}{\mu}\lVert \nabla f(w^t)\rVert + \mathbb{E}_k[\xi_k^t\lVert w^{t} -\tilde{w}_k^t \rVert].
\end{align}
Here, we further define  $\lVert \tilde{w}^t  - w^t\rVert$ as 
\begin{align}
	\notag \lVert \tilde{w}^t_k  - w^t\rVert  \leq 
    			 & \frac{1}{\mu}\lVert \nabla h(\tilde{w}^t_k,w^t) - \nabla h(w^t,w^t)\rVert\\
    \notag \leq &  \frac{1}{\mu}\lVert \nabla F_k(\tilde{w}^t_k) + \mu(\tilde{w}^t_k - w^t) \\
    \notag          &   - \nabla f(w^t) - \xi^t_k \mu (\tilde{w}^t_k - w^t) \rVert\\
   \notag 			\leq & \frac{1}{(1 - (1 -\xi^t_k))\mu} \lVert \nabla F_k(\tilde{w}^t_k) -  \nabla F_k(w^t) \rVert\\
   \leq &\frac{1}{\xi^t_k\mu}\lVert \nabla F_k(\tilde{w}^t_k) -  \nabla F_k(w^t)  \rVert.
\end{align}
According to $L$-smooth, we get the recursive inequality of $\lVert \tilde{w}^t  - w^t\rVert$ and $\lVert \nabla F_k(\tilde{w}^t) -  \nabla F_k(w^t) \rVert$ as follows:
\begin{align}
	\lVert \nabla F_k(\tilde{w}^t) - \nabla F_k(w^t) \rVert \leq L\lVert \tilde{w}^t - w^t \rVert,  \\
    \xi^t\lVert \tilde{w}^t - w^t \rVert \leq \frac{B}{\mu}\lVert \nabla f(\tilde{w}^t) - \nabla f(w^t) \rVert.
\end{align}
% Now we calculate the value of $\mathbb{E}_k[\xi_k^t]$. We set $g(t)=\sum_{n=1}^t\frac{p(1-p)^{n-1}}{n}+\frac{(1-p)^t}{t}$, which is a monotonically decreasing function. 
% \begin{align}
% 	      \max_t g(t) & = g(1) = p+(1-p)=1,\\
%     \notag \min_t g(t)   
%     &  = \lim_{t \to \infty}\sum_{n=1}^{t}\frac{p(1-p)^{n-1}}{n}\\
%     \notag & = \frac{p}{1-p}\sum_{n=1}^\infty\frac{(1-p)^n}{n}\\
%            & = \frac{p\ln{p}}{p-1}.
% \end{align}
% $k(p)=\frac{p\ln{p}}{p-1}=P$ is monotonically increasing, the value is in $(0, 1)$ when $p\in (0, 1)$.

Then, we get the inequality related to $f(\bar{w}^{t+1})$ and $f(w^t)$ based on $L$-smoothness, which can be given by:
\begin{align}
	\notag f(w^{t+1}) 
    \leq & f(w^t) + \langle \nabla f(w^t), \bar{w}^{t+1} - w^t \rangle 
                  + \frac{L}{2}\lVert\bar{w}^{t+1}-w^t\rVert^2\\
    \notag\leq & f(w^t) - \frac{1}{\mu}\lVert \nabla f(w^t)\rVert^2 + \frac{\gamma B}{\mu} \mathbb{E}_k[\langle \nabla f(w^t), e^t_k\rangle] \\
    \notag & - \frac{1}{\mu}\langle \nabla f(w^t), \mathbb{E}_k\left[ \nabla F_k(w^{t+1}) - \nabla F_k(w^t) \right]\rangle\\
    & \label{eq15} + \mathbb{E}_k \left[ \xi_k^t \langle \nabla f(w^t), \tilde{w}_k^t - w^{t+1} \rangle \right] + \frac{L}{2}\lVert w^{t+1} - w^t \rVert^2 
\end{align}
According to (8) in FedProx, the first three items are identical. So we only need to compare the last two items in (14) to $ \frac{L}{2}\lVert w^{t+1} - w^t \rVert^2$ at (8) in FedProx, which is equal to  $\frac{L}{2} \lVert \frac{1+\gamma}{\mu} \nabla F_k(w^t) \rVert^2$. 
Firstly, we analyze $\mathbb{E}_k \left[ \xi_k^t \langle \nabla f(w^t), \tilde{w}_k^t - w^{t+1} \rangle \right]$. 
\begin{align}
    \notag \mathbb{E}_k \left[ \xi_k^t \langle \nabla f(w^t), \tilde{w}_k^t - w^{t+1} \rangle \right] = & \mathbb{E}_k \left[ \xi_k^t \langle \nabla f(w^t), \tilde{w}_k^t - w^{t} \rangle \right] \\
    \notag & + \mathbb{E}_k \left[ \xi_k^t \langle \nabla f(w^t), w^t - w^{t+1} \rangle \right]
\end{align}
\begin{align}
    \notag \mathbb{E}_k \left[ \langle \nabla f(w^t), \tilde{w}_k^t - w^{t} \rangle \right] \leq & \mathbb{E}_k \left[ F_k(\tilde{w}_k^t) - F_k(w^{t})\right. \\
                      & \left. - \frac{\mu}{4}\lVert \tilde{w}_k^t - w^{t} \rVert^2 \right] \\
    \notag \mathbb{E}_k \left[ \langle \nabla f(w^t), w^t - w^{t+1} \rangle \right] \leq & \mathbb{E}_k \left[ F_k(w^t) - F_k(w^{t+1})\right. \\
    \notag            & - \frac{\mu}{4}\lVert w^t - w^{t+1} \rVert^2 \\
                      & \left. + L\lVert w^t - w^{t+1} \rVert^2 \right].
\end{align}
$ \frac{L}{2}\lVert w^{t+1} - w^t \rVert^2$ in FedTrip is
\begin{align}
     \notag \frac{L}{2}\lVert w^{t+1} - w^t \rVert^2 \leq & L \left[\lVert \frac{1+\gamma}{\mu} \nabla F_k(w^t) \rVert^2 \right. \\
        & + \left.\mathbb{E}_k \lVert\xi_k^t (\tilde{w}_k^t - w^t) \rVert^2\right].
\end{align}
Combine (15), (16) and (17), we have the additional items $Q^t$ in FedTrip: 
\begin{align}
    \notag Q^t = & \frac{L}{2} \lVert \frac{1+\gamma}{\mu} \nabla F_k(w^t) \rVert^2 + L \lVert\xi_k^t (\tilde{w}_k^t - w^t) \rVert^2 \\
    \notag & + \mathbb{E}_k \xi_k^t \left[  F_k(\tilde{w}_k^t) - F_k(w^{t+1}) - \frac{\mu}{4}\lVert \tilde{w}_k^t - w^{t} \rVert^2 \right. \\
           & - \frac{\mu}{4}\lVert w^t - w^{t+1} \rVert^2 + \left. L\lVert w^t - w^{t+1} \rVert^2 \right].
\end{align}
Now we analyze the items one by one. 
% \begin{align}
%     \notag \mathbb{E}_k \left[\lVert\xi_k^t (\tilde{w}_k^t - w^t) \rVert^2 + \lVert\xi_k^tL\lVert w^t - w^{t+1} \rVert^2\right] \leq \\
%     \notag \xi^t\lVert \tilde{w}^t - w^{t+1}\rVert^2 - \xi^t(1-\xi^t)L\lVert w^t - w^{t+1} \rVert^2,
% \end{align}
In FedProx, $\mu$ is set as $6LB^2$ as example, where $B\gg 1$. We adapt it and easily get $\left(\frac{\mu}{4} - L\right) \gg 0$ and $\left(\frac{\xi^t\mu}{4} - L\right) \gg 0$, where $\xi^t=\mathbb{E}[\xi_k^t] \in (0, 1]$.\\
Then we consider $\frac{L}{2} \lVert \frac{1+\gamma}{\mu} \nabla F_k(w^t) \rVert^2$. 
\begin{align}
    \mathbb{E}_k \left[\frac{L}{2} \lVert \frac{1+\gamma}{\mu} \nabla F_k(w^t) \rVert^2\right] = \frac{LB^2(1+\gamma)^2}{2\mu^2}\lVert \nabla f(w^t)\rVert^2.
\end{align}
From (11), we have $ - \frac{1+\gamma}{\mu}\lVert \nabla F_k(w^t) \rVert \leq \xi_k^t\lVert\tilde{w}_k^t - w^t\rVert - \lVert  w^{t+1} - w^{t} \rVert$.
\begin{align}
    \notag \frac{LB^2(1+\gamma)^2}{\mu^2}\lVert \nabla f(w^t)\rVert^2 \leq & L\left( (\xi_k^t)^2\lVert \tilde{w}_k^t - w^t \rVert^2 \right.\\
    \notag & - 2\xi_k^t \lVert \tilde{w}_k^t - w^t \rVert\lVert w^{t+1} - w^{t} \rVert \\
    \notag & \left. + \lVert w^{t+1} - w^{t} \rVert^2 \right)\\
    \notag \ll & \frac{\mu\xi_k^t}{4}\left(\lVert \tilde{w}_k^t - w^t \rVert^2\right.\\
    \notag & \left. + \lVert w^{t+1} - w^{t} \rVert^2 \right)
\end{align}
Lastly, we define $L_0$ as the local Lipschitz continuity constant of function $f$ and we have 
\begin{align}
    \notag \lVert f(\tilde{w}_k^t) - f(w^{t}) \rVert \leq L_0\lVert \tilde{w}_k^t - w^{t} \rVert
\end{align}
$L_0\lVert \tilde{w}_k^t - w^{t} \rVert - \frac{\mu}{4} \lVert \tilde{w}_k^t - w^{t} \rVert^2 < 0 $ is satisfied when $L_0 < \frac{\mu}{4} \lVert \tilde{w}_k^t - w^{t+1}\ \rVert\, \forall t$.

If $\mu, L, \gamma, \xi$ stasify, we have $\rho > 0 $ and $Q^t>0$.
Assume that $E_{S_t}[f(w^{t+1})] = f(w^{t+1)}$, (14) can be written as
\begin{align}
    \notag E_{S_t}[f(w^{t+1})] 
    \leq & f(w^t) + \left(\frac{1-\gamma B}{\mu} - \frac{L(1+\gamma)B}{\mu^2} \right. \\
         & \left.  -\frac{L(1+\gamma)^2B^2}{2\mu^2}\right)\lVert \nabla f(w^t) \rVert^2 - Q^t.
\end{align}
If we define that $\gamma=0$, the inequality is transformed to 
\begin{align}
	\notag f(w^{t+1}) \leq & f(w^t)  - \left(  \frac{1}{\mu} - \frac{LB}{\mu^2} - \frac{LB^2}{2\mu^2} \right) \lVert \nabla f(w^t) \rVert^2 - Q^t \\
    \leq & f(w^t) - \rho \lVert \nabla f(w^t) \rVert^2 -Q^t.
\end{align}

\section*{Acknowledgment}
This work was supported by the National Natural Science Foundation of China (No. 62072436) and the National Key Research and Development Program of China(2021YFB2900102). 

{\small
\bibliographystyle{IEEEtran}
\bibliography{ref}

\begin{thebibliography}{10}
\providecommand{\url}[1]{#1}
\csname url@samestyle\endcsname
\providecommand{\newblock}{\relax}
\providecommand{\bibinfo}[2]{#2}
\providecommand{\BIBentrySTDinterwordspacing}{\spaceskip=0pt\relax}
\providecommand{\BIBentryALTinterwordstretchfactor}{4}
\providecommand{\BIBentryALTinterwordspacing}{\spaceskip=\fontdimen2\font plus
\BIBentryALTinterwordstretchfactor\fontdimen3\font minus
  \fontdimen4\font\relax}
\providecommand{\BIBforeignlanguage}[2]{{%
\expandafter\ifx\csname l@#1\endcsname\relax
\typeout{** WARNING: IEEEtran.bst: No hyphenation pattern has been}%
\typeout{** loaded for the language `#1'. Using the pattern for}%
\typeout{** the default language instead.}%
\else
\language=\csname l@#1\endcsname
\fi
#2}}
\providecommand{\BIBdecl}{\relax}
\BIBdecl

\bibitem{he2016deep}
K.~He, X.~Zhang, S.~Ren \emph{et~al.}, ``Deep residual learning for image
  recognition,'' in \emph{CVPR}, 2016, pp. 770--778.

\bibitem{vaswani2017attention}
A.~Vaswani, N.~Shazeer, N.~Parmar \emph{et~al.}, ``Attention is all you need,''
  \emph{NIPS}, vol.~30, 2017.

\bibitem{devlin2018bert}
J.~Devlin, M.-W. Chang, K.~Lee \emph{et~al.}, ``Bert: Pre-training of deep
  bidirectional transformers for language understanding,'' \emph{arXiv preprint
  arXiv:1810.04805}, 2018.

\bibitem{liu2021swin}
Z.~Liu, Y.~Lin, Y.~Cao \emph{et~al.}, ``Swin transformer: Hierarchical vision
  transformer using shifted windows,'' in \emph{CVPR}, 2021, pp.
  10\,012--10\,022.

\bibitem{seneviratne2017survey}
S.~Seneviratne, Y.~Hu, T.~Nguyen \emph{et~al.}, ``A survey of wearable devices
  and challenges,'' \emph{IEEE Commun. Surv. Tutor.}, vol.~19, no.~4, pp.
  2573--2620, 2017.

\bibitem{zhang2019deep}
C.~Zhang, P.~Patras, and H.~Haddadi, ``Deep learning in mobile and wireless
  networking: A survey,'' \emph{IEEE Commun. Surv. Tutor.}, vol.~21, no.~3, pp.
  2224--2287, 2019.

\bibitem{lim2020federated}
W.~Y.~B. Lim, N.~C. Luong, D.~T. Hoang \emph{et~al.}, ``Federated learning in
  mobile edge networks: A comprehensive survey,'' \emph{IEEE Commun. Surv.
  Tutor.}, vol.~22, no.~3, pp. 2031--2063, 2020.

\bibitem{konevcny2016federated}
J.~Kone{\v{c}}n{\`y}, H.~B. McMahan, F.~X. Yu \emph{et~al.}, ``Federated
  learning: Strategies for improving communication efficiency,'' \emph{arXiv
  preprint arXiv:1610.05492}, 2016.

\bibitem{bonawitz2019towards}
K.~Bonawitz, H.~Eichner, W.~Grieskamp \emph{et~al.}, ``Towards federated
  learning at scale: System design,'' in \emph{MLSys}, vol.~1, 2019, pp.
  374--388.

\bibitem{yang2019federated}
Q.~Yang, Y.~Liu, Y.~Cheng \emph{et~al.}, ``Federated learning,'' \emph{Synth.
  Lect. Artif. Intell. Mach. Learn.}, vol.~13, no.~3, pp. 1--207, 2019.

\bibitem{wang2020federated}
H.~Wang, M.~Yurochkin, Y.~Sun \emph{et~al.}, ``Federated learning with matched
  averaging,'' in \emph{ICLR}, 2020.

\bibitem{wolfrath2022haccs}
J.~Wolfrath, N.~Sreekumar, D.~Kumar \emph{et~al.}, ``Haccs: Heterogeneity-aware
  clustered client selection for accelerated federated learning,'' in
  \emph{IPDPS}, 2022.

\bibitem{li2022federated}
Q.~Li, Y.~Diao, Q.~Chen \emph{et~al.}, ``Federated learning on non-iid data
  silos: An experimental study,'' in \emph{ICDE}, 2022, pp. 965--978.

\bibitem{mcmahan2017communication}
H.~B. McMahan, E.~Moore, D.~Ramage \emph{et~al.}, ``Communication-efficient
  learning of deep networks from decentralized data,'' in \emph{AISTATS}, 2017,
  pp. 1273--1282.

\bibitem{kaissis2020secure}
G.~A. Kaissis, M.~R. Makowski, D.~R{\"u}ckert \emph{et~al.}, ``Secure,
  privacy-preserving and federated machine learning in medical imaging,''
  \emph{Nat. Mach. Intell.}, vol.~2, no.~6, pp. 305--311, 2020.

\bibitem{khaled2020tighter}
A.~Khaled, K.~Mishchenko, and P.~Richt{\'a}rik, ``Tighter theory for local sgd
  on identical and heterogeneous data,'' in \emph{AISTATS}, 2020, pp.
  4519--4529.

\bibitem{karimireddy2020scaffold}
S.~P. Karimireddy, S.~Kale, M.~Mohri \emph{et~al.}, ``Scaffold: Stochastic
  controlled averaging for federated learning,'' in \emph{ICML}, 2020, pp.
  5132--5143.

\bibitem{li2020federated1}
T.~Li, A.~K. Sahu, A.~Talwalkar \emph{et~al.}, ``Federated learning:
  Challenges, methods, and future directions,'' \emph{IEEE Signal Process Mag},
  vol.~37, no.~3, pp. 50--60, 2020.

\bibitem{zhao2018federated}
Y.~Zhao, M.~Li, L.~Lai \emph{et~al.}, ``Federated learning with non-iid data,''
  \emph{arXiv preprint arXiv:1806.00582}, 2018.

\bibitem{hsieh2020non}
K.~Hsieh, A.~Phanishayee, O.~Mutlu, and P.~Gibbons, ``The non-iid data quagmire
  of decentralized machine learning,'' in \emph{ICML}, 2020, pp. 4387--4398.

\bibitem{kairouz2019advances}
P.~Kairouz, H.~B. McMahan, B.~Avent \emph{et~al.}, ``Advances and open problems
  in federated learning,'' \emph{arXiv preprint arXiv:1912.04977}, 2019.

\bibitem{wang2020tackling}
J.~Wang, Q.~Liu, H.~Liang \emph{et~al.}, ``Tackling the objective inconsistency
  problem in heterogeneous federated optimization,'' \emph{NIPS}, vol.~33, pp.
  7611--7623, 2020.

\bibitem{reddi2020adaptive}
S.~Reddi, Z.~Charles, M.~Zaheer \emph{et~al.}, ``Adaptive federated
  optimization,'' \emph{arXiv preprint arXiv:2003.00295}, 2020.

\bibitem{li2021fedbn}
X.~Li, M.~Jiang, X.~Zhang \emph{et~al.}, ``Fedbn: Federated learning on non-iid
  features via local batch normalization,'' \emph{arXiv preprint
  arXiv:2102.07623}, 2021.

\bibitem{li2020federated}
T.~Li, A.~K. Sahu, M.~Zaheer \emph{et~al.}, ``Federated optimization in
  heterogeneous networks,'' in \emph{MLSys}, vol.~2, 2020, pp. 429--450.

\bibitem{acar2021federated}
D.~A.~E. Acar, Y.~Zhao, R.~Matas \emph{et~al.}, ``Federated learning based on
  dynamic regularization,'' in \emph{ICLR}, 2021.

\bibitem{li2019feddane}
T.~Li, A.~K. Sahu, M.~Zaheer \emph{et~al.}, ``Feddane: A federated newton-type
  method,'' in \emph{ACSSC}, 2019, pp. 1227--1231.

\bibitem{kim2022communication}
G.~Kim, J.~Kim, and B.~Han, ``Communication-efficient federated learning with
  acceleration of global momentum,'' \emph{arXiv preprint arXiv:2201.03172},
  2022.

\bibitem{mendieta2022local}
M.~Mendieta, T.~Yang, P.~Wang \emph{et~al.}, ``Local learning matters:
  Rethinking data heterogeneity in federated learning,'' in \emph{CVPR}, 2022,
  pp. 8397--8406.

\bibitem{li2021model}
Q.~Li, B.~He, and D.~Song, ``Model-contrastive federated learning,'' in
  \emph{CVPR}, 2021, pp. 10\,713--10\,722.

\bibitem{schroff2015facenet}
F.~Schroff, D.~Kalenichenko, and J.~Philbin, ``Facenet: A unified embedding for
  face recognition and clustering,'' in \emph{CVPR}, 2015, pp. 815--823.

\bibitem{yao2021local}
D.~Yao, W.~Pan, Y.~Dai \emph{et~al.}, ``Local-global knowledge distillation in
  heterogeneous federated learning with non-iid data,'' \emph{arXiv preprint
  arXiv:2107.00051}, 2021.

\bibitem{hinton2015distilling}
G.~Hinton, O.~Vinyals, J.~Dean \emph{et~al.}, ``Distilling the knowledge in a
  neural network,'' \emph{arXiv preprint arXiv:1503.02531}, vol.~2, no.~7,
  2015.

\bibitem{hadsell2006dimensionality}
R.~Hadsell, S.~Chopra, and Y.~LeCun, ``Dimensionality reduction by learning an
  invariant mapping,'' in \emph{CVPR}, vol.~2, 2006, pp. 1735--1742.

\bibitem{karimireddy2020mime}
S.~P. Karimireddy, M.~Jaggi, S.~Kale \emph{et~al.}, ``Mime: Mimicking
  centralized stochastic algorithms in federated learning,'' \emph{arXiv
  preprint arXiv:2008.03606}, 2020.

\bibitem{lecun1998gradient}
Y.~LeCun, L.~Bottou, Y.~Bengio \emph{et~al.}, ``Gradient-based learning applied
  to document recognition,'' \emph{Proc. IEEE}, vol.~86, no.~11, pp.
  2278--2324, 1998.

\bibitem{van2008visualizing}
L.~Van~der Maaten and G.~Hinton, ``Visualizing data using t-sne.'' \emph{J Mach
  Learn Res}, vol.~9, no.~11, 2008.

\bibitem{paszke2019pytorch}
A.~Paszke, S.~Gross, F.~Massa \emph{et~al.}, ``Pytorch: An imperative style,
  high-performance deep learning library,'' \emph{NIPS}, vol.~32, pp.
  8026--8037, 2019.

\bibitem{caldas2018leaf}
S.~Caldas, S.~M.~K. Duddu, P.~Wu, T.~Li, J.~Kone{\v{c}}n{\`y}, H.~B. McMahan,
  V.~Smith, and A.~Talwalkar, ``Leaf: A benchmark for federated settings,''
  \emph{arXiv preprint arXiv:1812.01097}, 2018.

\bibitem{xiao2017fashion}
H.~Xiao, K.~Rasul, and R.~Vollgraf, ``Fashion-mnist: a novel image dataset for
  benchmarking machine learning algorithms,'' \emph{arXiv preprint
  arXiv:1708.07747}, 2017.

\bibitem{cohen2017emnist}
G.~Cohen, S.~Afshar, J.~Tapson \emph{et~al.}, ``Emnist: Extending mnist to
  handwritten letters,'' in \emph{IJCNN}, 2017, pp. 2921--2926.

\bibitem{krizhevsky2010cifar}
A.~Krizhevsky, V.~Nair, and G.~Hinton, ``Cifar-10 (canadian institute for
  advanced research),'' \emph{URL http://www. cs. toronto. edu/kriz/cifar.
  html}, vol.~5, no.~4, p.~1, 2010.

\bibitem{nair2010rectified}
V.~Nair and G.~E. Hinton, ``Rectified linear units improve restricted boltzmann
  machines,'' in \emph{ICML}, 2010.

\bibitem{krizhevsky2012imagenet}
A.~Krizhevsky, I.~Sutskever, and G.~E. Hinton, ``Imagenet classification with
  deep convolutional neural networks,'' \emph{NIPS}, vol.~25, 2012.

\bibitem{wang2019slowmo}
J.~Wang, V.~Tantia, N.~Ballas \emph{et~al.}, ``Slowmo: Improving
  communication-efficient distributed sgd with slow momentum,'' \emph{arXiv
  preprint arXiv:1910.00643}, 2019.

\bibitem{sutskever2013importance}
I.~Sutskever, J.~Martens, G.~Dahl \emph{et~al.}, ``On the importance of
  initialization and momentum in deep learning,'' in \emph{ICML}, 2013, pp.
  1139--1147.

\end{thebibliography}
}

\end{document}